\documentclass[final,1p,times]{elsarticle}
\usepackage{cases}
\usepackage{amssymb}
\usepackage{amsthm}

\newcommand{\ds}{\displaystyle}

\journal{Wave Motion}

\begin{document}



\begin{frontmatter}

\title{Fast and slow dynamics in a nonlinear elastic bar excited by longitudinal vibrations}

\author[IT]{Nicolas Favrie}
\ead{nicolas.favrie@univ-amu.fr}
\author[LMA]{Bruno Lombard}
\ead{lombard@lma.cnrs-mrs.fr}
\author[LMA]{C\'edric Payan}
\ead{cedric.payan@univ-amu.fr}
\cortext[cor1]{Corresponding author. Tel.: +33 491 16 44 13.}
\address[IT]{IUSTI, Aix-Marseille Universit\'e, UMR CNRS 7343, 5 rue E. Fermi, 13453 Marseille Cedex 13, France}
\address[LMA]{Laboratoire de M\'{e}canique et d'Acoustique, UPR 7051 CNRS, 31 chemin Joseph Aiguier, 13402 Marseille, France}

\begin{abstract}
The dynamics of heterogeneous materials, like rocks and concrete, is complex. It includes such features as nonlinear elasticity, hysteresis, and long-time relaxation. This dynamics is very sensitive to microstructural changes and damage. The goal of this paper is to propose a physical model describing the longitudinal vibrations in heterogeneous material, and to develop a numerical strategy to solve the evolution equations. The theory relies on the coupling of two processes with radically different time scales: a fast process at the frequency of the excitation, governed by nonlinear elasticity and viscoelasticity, and a slow process, governed by the evolution of defects. The evolution equations are written as a nonlinear hyperbolic system with relaxation. A time-domain numerical scheme is developed, based on a splitting strategy. The features observed by numerical simulations show qualitative agreement with the features observed experimentally by Dynamic Acousto-Elastic Testing.
\end{abstract}

\begin{keyword}
Nonlinear acoustics; time-dependent materials; viscoelasticity; acoustic conditioning; numerical methods; hyperbolic system.
\end{keyword}

\end{frontmatter}


\section{Introduction}\label{SecIntro}

Understanding the mechanisms of acoustic nonlinearity in heterogeneous materials is an object of intensive studies \cite{Guyer99,Ostrovsky01,Guyer09,Lebedev14}. Experimental evidence has shown that media such as rocks and concrete possess an anomalously strong acoustic nonlinearity, which is of great importance for the description of ultrasonic phenomena including damage diagnostics. Besides the widely-studied nonlinear and hysteretic stress-strain relation \cite{Kadish96}, a long-time relaxation is also reported by most of the authors \cite{TenCate96,TenCate00}. This slow dynamics is typically observed in experiments of softening / hardening \cite{Renaud12,Riviere13}, where a bar is forced by a monochromatic excitation on a time interval, before the source is switched-off. During the experiment, the elastic modulus is measured by Dynamic Acousto-Elastic Testing methods. It can be observed that the elastic modulus decreases gradually (softening), and then it recovers progressively its initial value after the extinction of the source (hardening). The time scales of each stage is much longer than the time scale of the forcing, which justifies the term "slow dynamics".

The modelling of this slow dynamic effect has been investigated by many authors. An essentially phenomenological model is widely used for this purpose: the Preisach-Mayergoyz model (P-M model) based on the integral action of hysteretic elements connecting stress and strain \cite{Scalerandi06,Scalerandi10,Lebedev14}. This model initially arose from the theory of magnetism, where the "hysteron" has a clear physical significance. In elasticity, such a physical interpretation is not available. To overcome this limitation and to develop a rigorous theory, various authors have proposed alternative models based on clear mechanical concepts. To our knowledge, the first physical model of slow dynamics was described in \cite{TenCate00}, where the relaxation was related to the recovery of microscopic contact impeded by a smooth spectrum of energy barriers. This theory was extended in \cite{Aleshin07a,Aleshin07b}, and recently improved based on the analysis of inter-grain contacts and the resulting surface force potential with a barrier \cite{Lebedev14}. Another approach was followed in \cite{Pecorari04}, where the author shows that two rough surfaces interacting via adhesion forces yield dynamics similar to that of the fictitious elements of the Preisach-Mayergoyz space \cite{Pecorari04}.

Here, we present an alternative mechanical description of slow dynamics based on the works of Vakhnenko and coauthors \cite{Vakhnenko04,Vakhnenko05}, where the following scenario is proposed:
\begin{itemize}
\item the Young's modulus $E$ varies with time. One can write $E(g)$, where $g$ is a time-dependent concentration of defects. It is closely related to the notion of damage in solids mechanics. But contrary to what happens in this irreversible case, where $g$ strictly increases with time, the evolution of $g$ is reversible. Waiting a sufficiently long time, the initial material properties are recovered;
\item at equilibrium, stress $\sigma$ yields a concentration of defects $g_\sigma$. The dependence of $g_\sigma$ with respect to $\sigma$ is monotonic;
\item out of equilibrium, relaxation times are required for $g$ to reach $g_\sigma$. Whether $g<g_\sigma$ (increase in the number of defects) or $g>g_\sigma$ (decrease in the number of defects), Vakhnenko et al state that the time scales differ. The argument is given in section III of \cite{Vakhnenko05}: "there are various ways for an already existing crack in equilibrium to be further expanded when surplus tensile load is applied. However, under compressive load a crack, once formed, has only one spatial way to be annhilated or contracted". In both cases, these relaxation times are much longer than the time scale of the excitation, which explains the slow dynamics.
\end{itemize}
Comparisons with experimental data are given in section V of \cite{Vakhnenko05}, where the authors reproduced experiments done on Berea sandstone \cite{TenCate96}. One current weakness is that no micro-mechanical description of the involved defects has been proposed so far. A possible analogy may be found with populations of open / closed cracks filled with air, equivalent to a population of bubbles that relax towards an equilibrium state, depending on the applied stress \cite{Emelianov04,Gavrilyuk04}. In counterpart, one attractive feature of Vakhnenko's model is that it combines hyperbolic equations and relaxation terms, which constitutes a sound basis of physical phenomena \cite{Godunov03}. 

The present paper is a contribution to the theoretical analysis of this model and to its practical implementation to describe wave motion in damaged media. First, we point out that no mechanisms prevents the concentration of defects from exceeding 1, which is physically unrealistic. We fix this problem by proposing another expression for the equilibrium concentration. Second, the Stokes model describing viscoelasticity behaviour in \cite{Vakhnenko05} poorly describes the attenuation in real media, and it is badly suited to time-domain simulations of wave propagation. Instead, we propose a new nonlinear version of the Zener model. This viscoelastic model degenerates correctly towards a pure nonlinear elasticity model when attenuation effects vanish. Moreover, the usual Zener model in the linear regime is recovered \cite{Carcione07}. In practice, this model only requires one physical parameter under the assumption of constant quality factor. Third, hyperbolicity is analyzed. Depending on the chosen model of nonlinear elasticity, a real sound speed may obtained only on a finite interval of strains; this is true in  particular with the widely-used Landau's model.

The main effort of Vakhnenko et al was devoted to the construction of a model of slow dynamics. The resolution of the involved equations was quite rudimentary and not satisfying. Indeed, the equilibrium concentration of defects $g_\sigma$ was assumed to be known and was imposed (eq (17) in \cite{Vakhnenko05}), while it depends on $\sigma$. But treating the full coupled nonlinear equations is out of reach of a semi-analytical approach, which explains the strategy of these authors. On the contrary, we propose here a numerical method to integrate the full system of equations, involving the nonlinear elasticity, the hysteretic terms of viscoelasticity, and the slow dynamics. Due to the existence of different time scales, a splitting strategy is followed, ensuring the optimal time step for integration. The full system is split into a propagative hyperbolic part (resolved by a standard scheme for conservation laws) and into a relaxed part (resolved exactly). 

Our numerical model is very modular. The various bricks (nonlinear elasticity, viscoelasticity, slow dynamics) can be incorporated easily. Numerical tests validate each part separately. When all the whole bricks are put together, typical features of wave motion in damaged media are observed. The softening / hardening experiments are qualitatively reproduced. 


\section{Physical modeling}\label{SecPhys}

In this section, we write the basic components describing the wave motion in a 1D material with damage. The fundations rely on linear elastodynamics, whose equations are recalled in section \ref{SecPhysLin}. Then, the soft-ratchet model of Vakhnenko and coauthors is introduced and enhanced in section \ref{SecPhysSoft}. The fast dynamics is described in section \ref{SecPhysElasto}, where various known models of nonlinear elasticity are presented, and a nonlinear model of viscoelasticity is proposed. This latter degenerates correctly in the limit cases of linear elasticity or null attenuation.

\subsection{Linear elastodynamics}\label{SecPhysLin}

In the case of small deformations, the propagation of 1D elastic waves can be described by the following system \cite{Achenbach73}:
\begin{subnumcases}{\label{ElastoLin}}
\frac{\partial v}{\partial t}-\frac{1}{\rho}\frac{\partial \sigma}{\partial x}=\gamma,\label{ElastoLin1}\\
[6pt]
\frac{\partial \varepsilon}{\partial t}-\frac{\partial v}{\partial x}=0,\label{ElastoLin2}
\end{subnumcases}
where $t$ is the time, $x$ is the spatial coordinate, $\gamma$ is a forcing term, $u$ is the displacement, $v=\frac{\partial u}{\partial t}$ is the velocity, $\varepsilon=\frac{\partial u}{\partial x}$ is the strain, and $\sigma$ is the stress. The latter is a function of strain: $\sigma=\sigma(\varepsilon)$. 

In the linear case, Hooke's law writes $\sigma=E\,\varepsilon$, where $E$ is the Young's modulus, which is assumed to be constant over time. In the particular case where $\gamma$ is a Dirac source at $x_s$ with time evolution ${\cal G}(t)$, then the exact solution of (\ref{ElastoLin}) is straightforward
\begin{equation}
\varepsilon=-\frac{\mbox{sgn}(x-x_s)}{2\,c^2}\,{\cal G}\left(t-\frac{|x-x_s|}{c}\right),
\label{ElastoExact}
\end{equation}
where sgn is the sign distribution, and $c=\sqrt{\frac{1}{\rho}\frac{\partial \sigma}{\partial \varepsilon}}\equiv\sqrt{E/\rho}$ is the speed of sound.

The goal of the forthcoming sections is to extend the model (\ref{ElastoLin}) in three ways:
\begin{itemize}
\item time variations of $E$ due to the stress;
\item nonlinear Hooke's law;
\item hereditary effects (viscoelasticity).
\end{itemize}
The time scales for the first effect (variation of $E$) are much greater than for the second and third effect. This is consequently referred to as {\it slow dynamics}.


\subsection{Slow dynamics: soft-ratchet model}\label{SecPhysSoft}

\begin{figure}[h!]
\begin{center}
\begin{tabular}{cc}
(a) & (b)\\
\hspace{-0.9cm}
\includegraphics[scale=0.42]{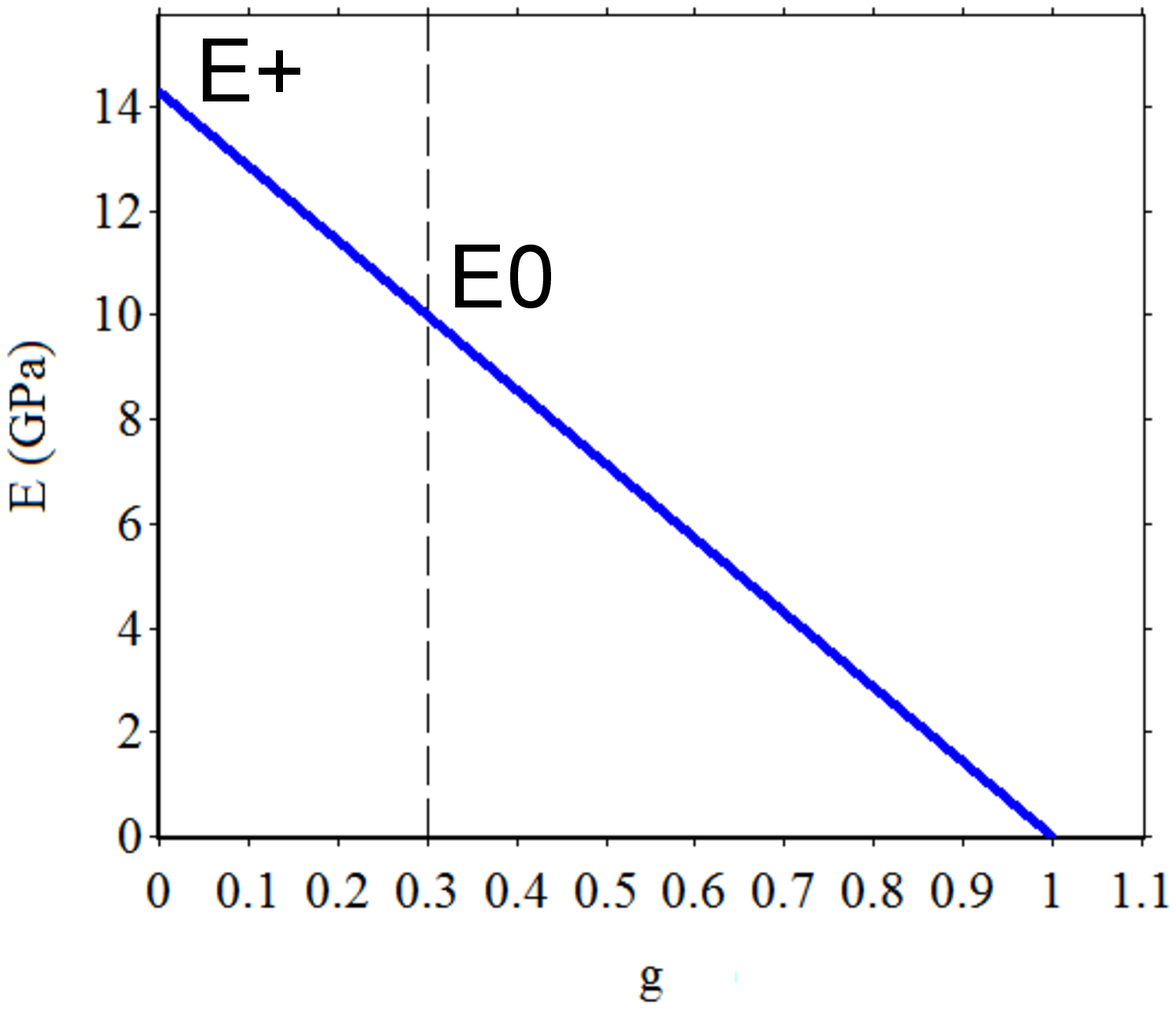} &
\hspace{-0.4cm}
\includegraphics[scale=0.42]{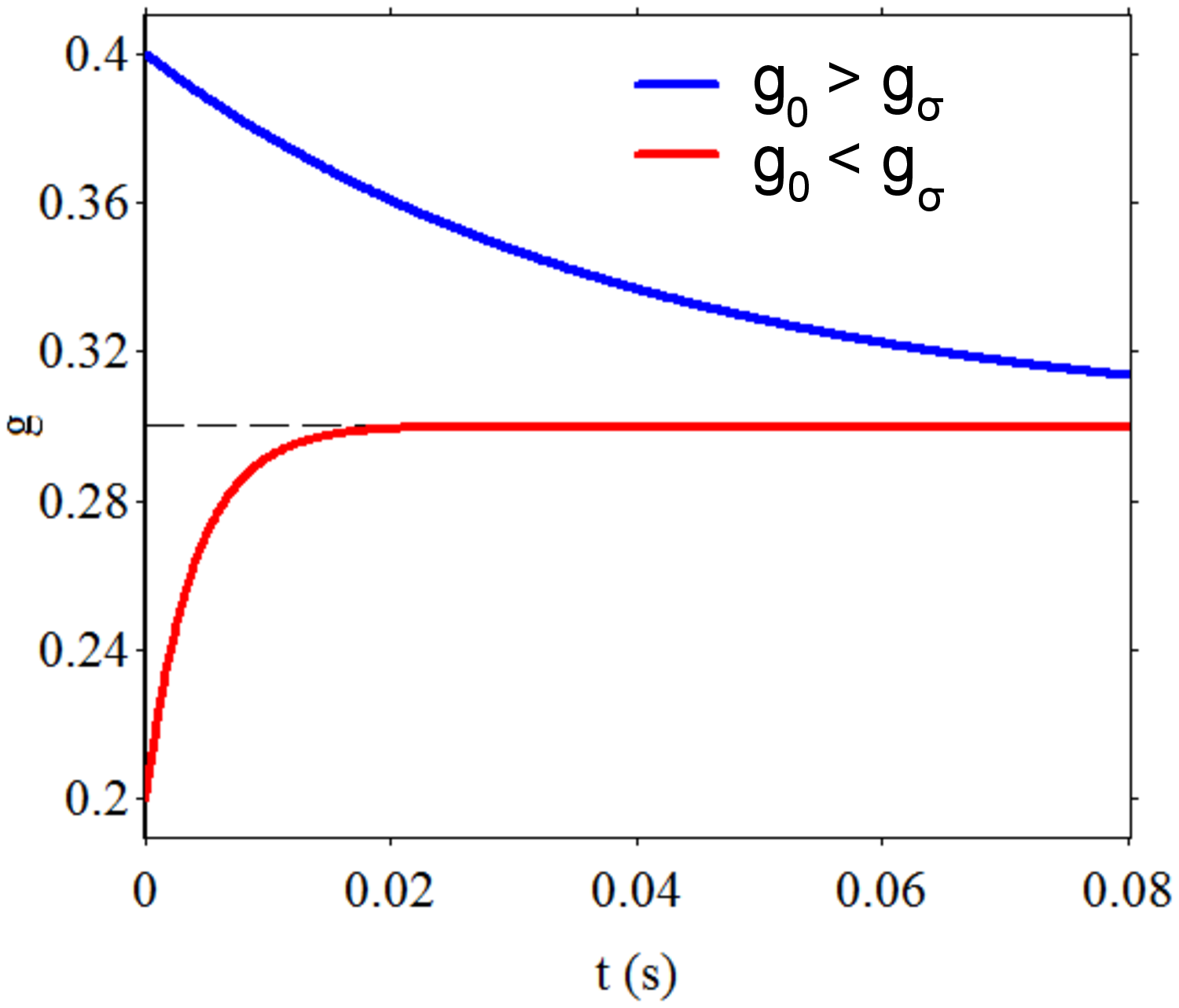} 
\end{tabular}
\caption{\label{FigYoung} parameters of the slow dynamics. (a): Young's modulus $E$ in terms of the concentration of defects $g$ (\ref{Young}), for $E^+=14.28$ GPa; the vertical dotted line denotes the initial concentration of defects $g_0=0.3$ and the corresponding Young's modulus $E_0=E(g_0)=10$ GPa. (b): time evolution of the concentration of defects $g$ given an equilibrium stress $\sigma$ and two initial values $g_0$; the horizontal dotted line denotes $g_\sigma$.}
\end{center}
\end{figure}

Here we follow the approach taken from \cite{Vakhnenko04,Vakhnenko05} with some modifications. The slow dynamics of the medium is assumed to rely on the concentration of activated defects $g$, which varies with $\sigma$. In the lowest approximation, the Young's modulus is written:
\begin{equation}
E=\left(1-\frac{\textstyle g}{\textstyle g_{cr}}\right)\,E^+,
\label{Young}
\end{equation}
where $g_{cr}$ and $E^+$ are the critical concentration of defects and the maximum possible value of Young's modulus, respectively (figure \ref{FigYoung}-(a)). The following constraints hold:
\begin{equation}
0 \leq g\leq g_{cr}\leq 1.
\label{HypVakh}
\end{equation} 

The concentration $g$ is assumed to evolve to its stress-dependent equilibrium value $g_\sigma$ at a rate $f_r$ if $g>g_\sigma$ (restoration), or $f_d$ if $g<g_\sigma$ (destruction). This mechanism can be modeled by the ordinary differential equation
\begin{equation}
\frac{\textstyle dg}{\textstyle dt}=-\left(f_r\,H(g-g_\sigma)+f_d\,H(g_\sigma-g)\right)\,(g-g_\sigma),
\label{Kinetic}
\end{equation}
where $H$ is the Heaviside step distribution. The frequencies $f_r$ and $f_d$ differ substantially: 
\begin{equation}
f_r \ll f_d \ll f_c,
\label{TimeScales}
\end{equation} 
where $f_c$ is a typical frequency of the excitation. Figure \ref{FigYoung}-(b) represents the time evolution of $g$, given a constant equilibrium concentration $g_\sigma=0.3$ denoted by a horizontal dotted line. The restoration and rupturation frequencies are $f_r=25$ Hz and $f_d=250$ Hz, respectively. Two initial value of the concentration of defects are considered: $g_0=0.2$ and $g_0=0.4$. In both cases, $g$ tends towards $g_\sigma$ with different rates: destruction is much faster than restoration. 

\begin{figure}[htbp]
\begin{center}
\begin{tabular}{cc}
(a) & (b)\\
\hspace{-0.9cm}
\includegraphics[scale=0.33]{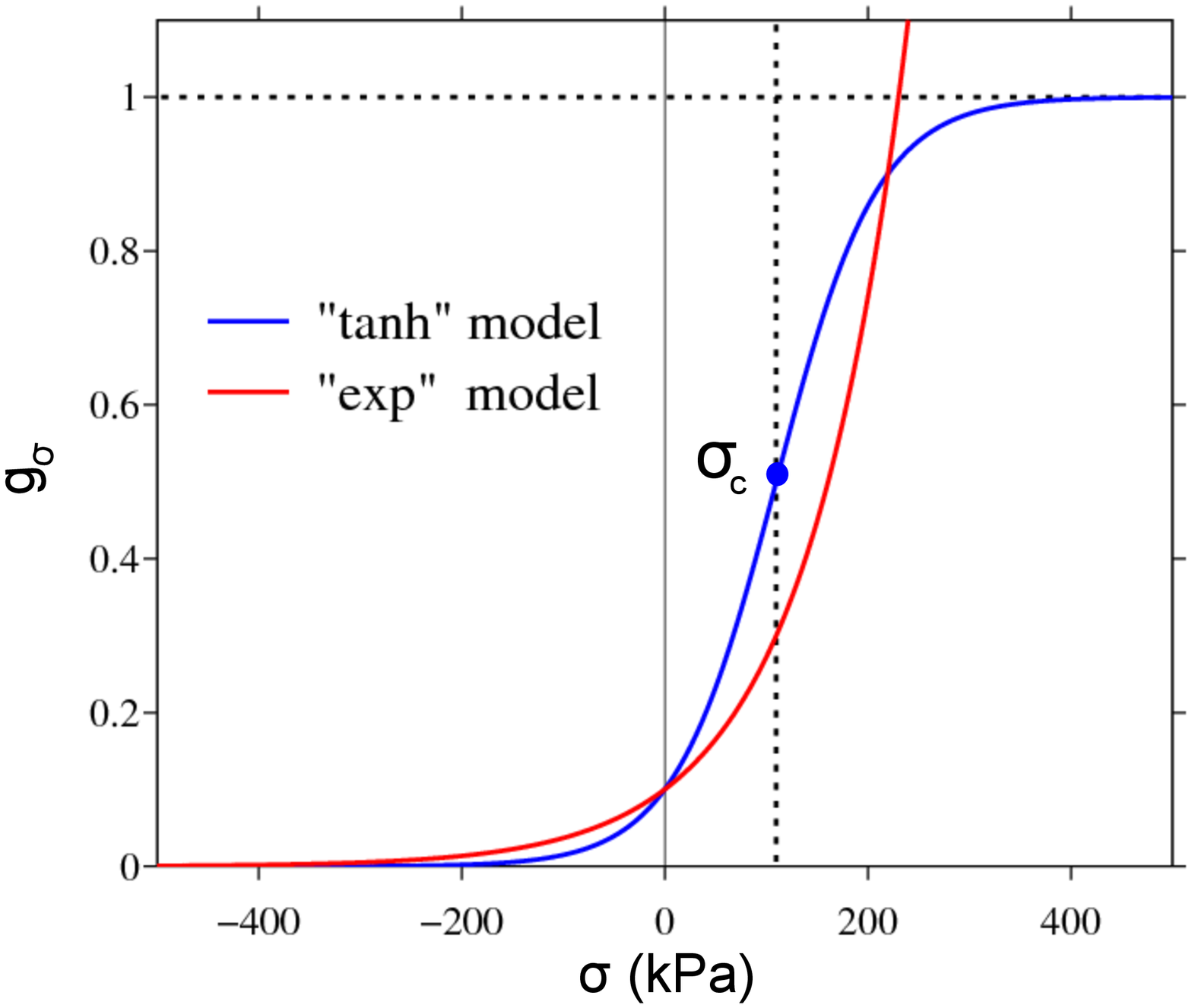} &
\hspace{-0.5cm}
\includegraphics[scale=0.33]{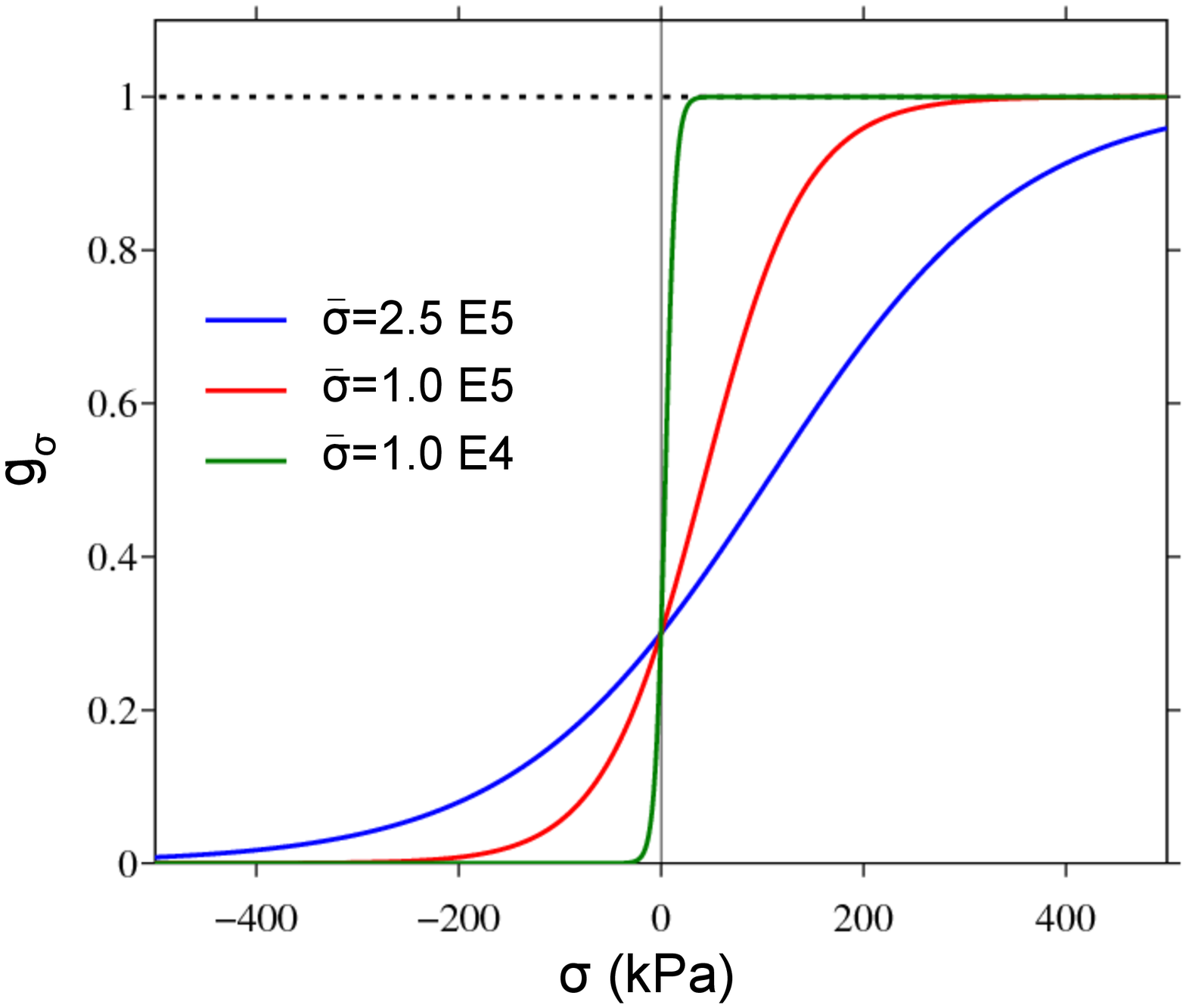}
\end{tabular}
\caption{\label{FigSRsigbar} equilibrium concentration of defects $g_\sigma$ in terms of the applied stress $\sigma$. (a): "exponential model" (\ref{GsigVakh}) and "tanh model" (\ref{GsigFLP}). (b): "tanh model" (\ref{GsigFLP}) with various values of $\overline{\sigma}$. The horizontal dotted line denotes the critical concentration of defects $g_{cr}$; the vertical dotted line denotes the central stress $\sigma_c$. }
\end{center}
\end{figure}

It remains to define the evolution of $g_\sigma$ with $\sigma$. In \cite{Vakhnenko04,Vakhnenko05}, the authors propose the expression
\begin{equation}
g_\sigma=g_0\,\exp(\sigma/\overline{\sigma}), \hspace{1cm}\overline{\sigma}=\frac{\textstyle kT}{\textstyle \upsilon},
\label{GsigVakh}
\end{equation}
where $g_0$ is the unstrained equilibrium concentration of defects, $k$ is the Boltzmann constant, $T$ is the temperature, and $\upsilon$ is a typical volume accounting for a single defect. If $ \sigma>\overline{\sigma}\ln g_{cr}/g_0$, then $g_\sigma>g_{cr}$; in this case, the concentration may evolve to $g>g_{cr}$ due to equation (\ref{Kinetic}), which contradicts the second assumption in (\ref{HypVakh}). To remove this drawback and to build a physically realistic expression of $g_\sigma$, we enforce (\ref{HypVakh}) together with the following requirements: 
\begin{subnumcases}{\label{Hyp}}
0\leq g_\sigma<g_{cr},\label{Hyp1}\\
g_\sigma(0)=g_0,\label{Hyp3}\\
\lim_{\sigma\rightarrow -\infty}g_\sigma=0,\label{Hyp2}\\
\lim_{\sigma\rightarrow +\infty}g_\sigma=g_{cr},\label{Hyp4}\\
\frac{\textstyle \partial g_\sigma}{\textstyle \partial \sigma}>0.\label{Hyp5}
\end{subnumcases}
The simplest smooth function satisfying (\ref{Hyp}) is
\begin{equation}
g_\sigma=\frac{\textstyle g_{cr}}{\textstyle 2}\left(1+\tanh\left(\frac{\textstyle \sigma-\sigma_c}{\textstyle \overline{\sigma}}\right)\right),
\label{GsigFLP}
\end{equation}
where the central stress is
\begin{equation}
\sigma_c=\overline{\sigma}\,\tanh^{-1}\left(1-2\frac{\textstyle g_0}{\textstyle g_{cr}}\right).
\label{Sig0}
\end{equation}
Figure \ref{FigSRsigbar}-(a) illustrates the two expressions of the stress-dependent equilibrium value $g_\sigma$: the "exponential model" (\ref{GsigVakh}), and the "tanh model" (\ref{GsigFLP})-(\ref{Sig0}). The numerical values are $g_0=0.3$ and $\overline{\sigma}=10^5$ Pa. The two expressions are the same at null stress. But for tractions greater than 230 kPa, the value of $g_\sigma$ deduced from (\ref{GsigVakh}) exceeds 1, leading to non-physical negative Young's modulus. Figure \ref{FigSRsigbar}-(b) illustrates the influence of $\overline{\sigma}$ in (\ref{GsigFLP}). As $\overline{\sigma}$ decreases, $g_\sigma$ may evolve more easily towards the extreme values 0 and $g_{cr}$, and hence the damage may increase thanks to (\ref{Kinetic}).

\subsection{Fast dynamics: nonlinear viscoelasticity}\label{SecPhysElasto}

\begin{figure}[htbp]
\begin{center}
\begin{tabular}{cc}
(a) & (b)\\
\hspace{-0.9cm}
\includegraphics[scale=0.35]{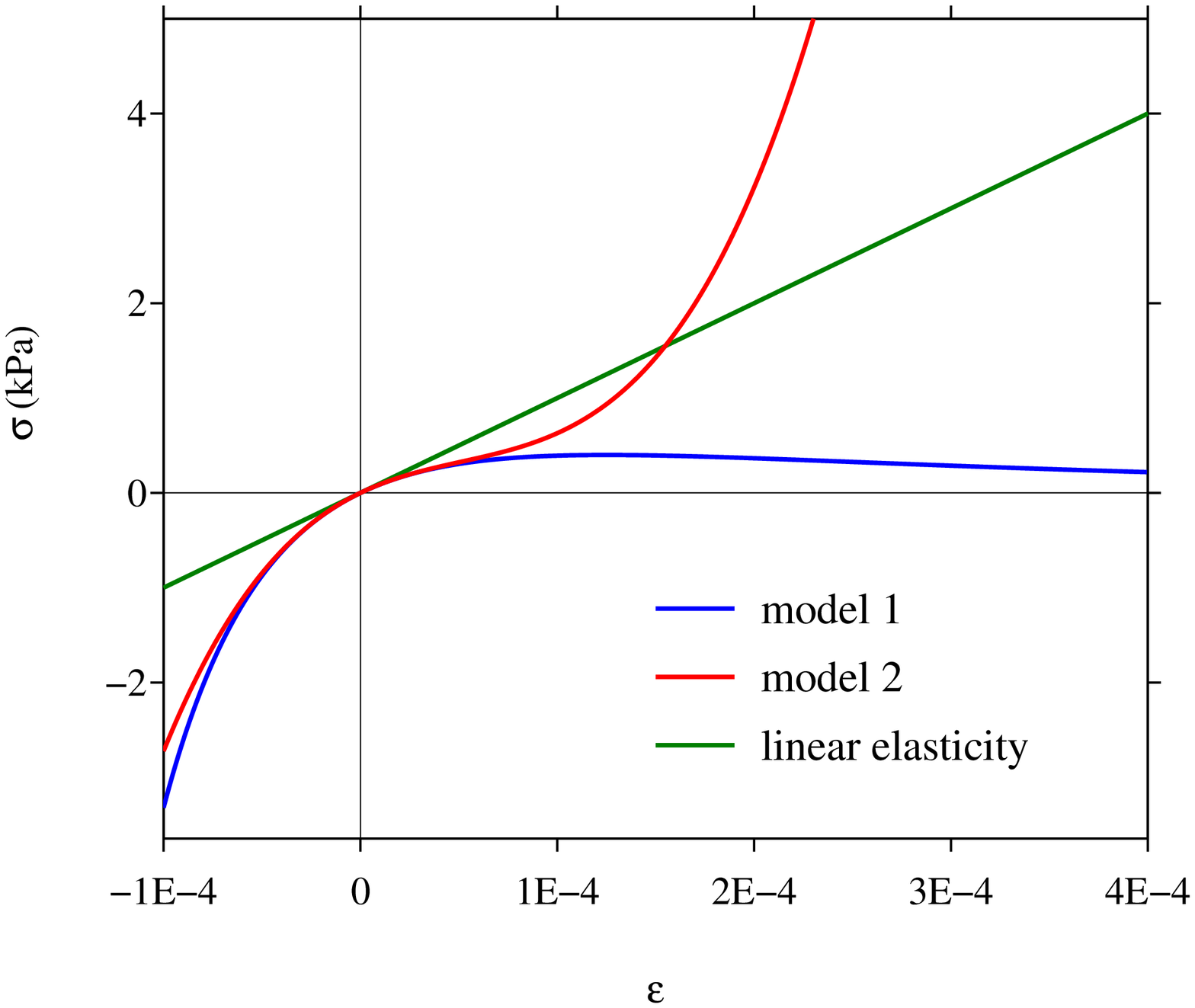} & 
\hspace{-0.9cm}
\includegraphics[scale=0.35]{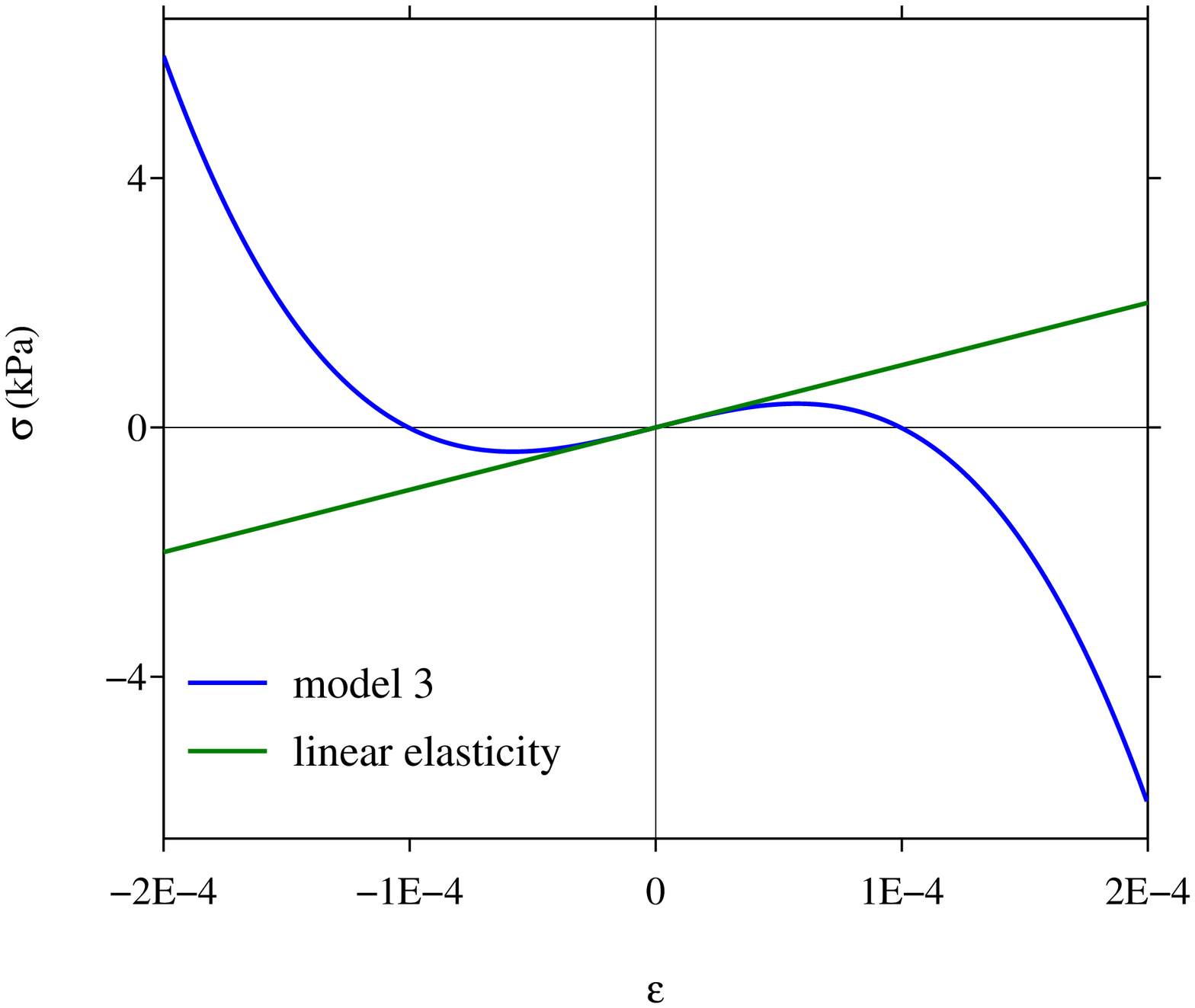} 
\end{tabular}
\caption{\label{FigElastoNL} Stress-strain relations for the three models (\ref{ElastoNL}). In (a), the dotted lines denote the coordinates of the inflexion point for model 1. The physical parameters are: $E=10$ GPa, $d=4.3\,10^{-4}$ m, $a=2$, $r=4$ (models 1 and 2), $\beta=100$, $\delta=10^8$ (model 3).}
\end{center}
\end{figure}

\paragraph{Nonlinear elasticity} ~\\
The stress-strain relation is given by a smooth function
\begin{equation}
s\equiv s(\epsilon,\,K,\,{\bf p}),
\label{ElastoNL}
\end{equation}
where $s$ is the stress, $\epsilon$ is the strain, $K$ is a stiffness, and ${\bf p}$ is a set of parameters governing the nonlinearity. No pre-stress is considered; $K$ is the slope of $s$ at the origin; lastly, $s$ is homogeneous of degree 1 in $K$. In other words, $s$ satisfies the following properties:
\begin{equation}
s(0,\,K,\,{\bf p})=0,\hspace{0.5cm} \frac{\partial s}{\partial \epsilon}(0,\,K,\,{\bf p})=K,\hspace{0.5cm} s(\epsilon,\,\alpha K,{\bf p})=\alpha\,s(\epsilon,\,K,\,{\bf p}).
\label{PropElastoNL}
\end{equation}
Three models of nonlinear elasticity (\ref{ElastoNL}) satisfying (\ref{PropElastoNL}) are now given and illustrated in figure \ref{FigElastoNL}.

\noindent
\underline{Model 1}. This model is from \cite{Vakhnenko05} and mimics the Lennard-Jones potential describing the interaction between a pair of neutral atoms:
\begin{equation}
s(\epsilon,\,K,\,{\bf p})=K \frac{\textstyle d}{\textstyle r-a}\left(\frac{1}{\left(\displaystyle 1+\frac{\epsilon}{d}\right)^{a+1}}-\frac{1}{\left(\displaystyle 1+\frac{\epsilon}{d}\right)^{r+1}}\right),\hspace{0.5cm} {\bf p}=(r,\,a,\,d)^T.
\label{Model1}
\end{equation}
The nonlinear parameters are the repulsion and attraction coefficients $r$ and $a$ ($0<a<r$). The strain is bounded below by the maximal allowable closure $d$. The function (\ref{Model1}) has an extremal point $\epsilon_c>0$, and then it decreases asymptotically towards 0 when $\epsilon>\epsilon_c$ (figure \ref{FigElastoNL}-(a)).

\noindent
\underline{Model 2}. A third-order Taylor expansion of the model 1 (\ref{Model1}) yields
\begin{equation}
s(\epsilon,\,K,\,{\bf p})=K\,\epsilon\left(1-\frac{1}{2}\left(r+a+3\right)\frac{\textstyle \epsilon}{\textstyle d}+\frac{\textstyle 1}{\textstyle 6}\left(r^2+ra+a^2+6r+6a+11\right)\left(\frac{\textstyle \epsilon}{\textstyle d}\right)^2\right),\hspace{0.5cm}{\bf p}=(r,\,a,\,d)^T.
\label{Model2}
\end{equation}
The nonlinear parameters are the same than in model 1. But contrary to what happened in model 1, the function (\ref{Model2}) is a strictly monotonically increasing function without extremal point (figure \ref{FigElastoNL}-(a)). Moreover, the strain is not bounded below.

\noindent
\underline{Model 3}. The most widely used law in ultrasonic NonDestructive Testing is the so-called Landau's model \cite{Landau70}
\begin{equation}
s(\epsilon,\,K,\,{\bf p})=K\,\epsilon\left(1-\beta\,\epsilon-\delta\,\epsilon^2\right),\hspace{0.5cm} {\bf p}=(\beta,\,\delta)^T.
\label{Model3}
\end{equation}
The parameters governing the nonlinear behavior are $\beta$ and $\delta$; in practice, $\beta \ll \delta$. Like what happens with model 1, the function (\ref{Model3}) has extremal points, but it is not bounded below (figure \ref{FigElastoNL}-(b)).

\begin{figure}[htbp]
\begin{center}
\begin{tabular}{cc}
\includegraphics[scale=0.30]{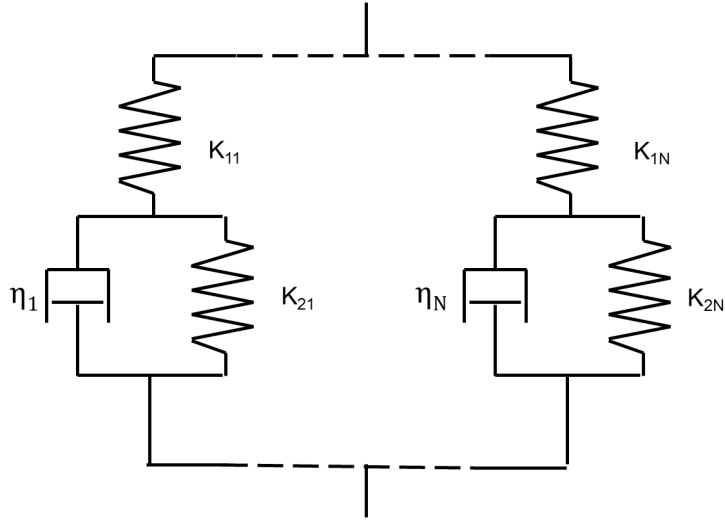} 
\end{tabular}
\caption{\label{FigZener} Rheological model of a generalized Zener material.}
\end{center}
\end{figure}

\paragraph{Viscoelasticity} ~\\
To incorporate attenuation, the following criteria are used as a guideline:
\begin{description}
\item[${\cal C}_1$:] when the viscous effects are null, the nonlinear elasticity must be recovered (\ref{ElastoNL});
\item[${\cal C}_2$:] when a linear stress-strain relation holds, it is necessary to recover the standard linear solid model (or generalized Zener model), which accurately represents the behavior of usual solids \cite{Carcione07}.
\end{description}
For this purpose, a system with $N$ Zener elements connected in parallel is considered (figure \ref{FigZener}). The total stress acting on the system is 
\begin{equation}
\sigma=\sum_{\ell=1}^N\sigma_{1\ell}=\sum_{\ell=1}^N(\sigma_{2\ell}+\sigma_{3\ell}),
\label{MailleSigma}
\end{equation}
where the index 1 refers to the springs in series, and indices 2-3 refer to the springs and dashpots in parallel. The strain $\varepsilon$ is
\begin{equation}
\varepsilon=\varepsilon_{1\ell}+\varepsilon_{2\ell},\hspace{1cm}\ell=1,\cdots,N.
\label{MailleEps}
\end{equation}
The index 1 springs satisfy nonlinear stress-strain relations (\ref{ElastoNL}) with stiffnesses $K_{1\ell}$. The parameters ${\bf p}$ governing the nonlinearity (for instance $\beta$ and $\delta$ in model 3 (\ref{Model3})) are assumed to be constant and identical for each element. The index 2 springs satisfy linear stress-strain relations with stiffnesses $K_{2\ell}$. Lastly, the dashpots satisfy linear Maxwell laws with coefficients of viscosity $\eta_\ell$. These laws are summed up as follows:
\begin{subnumcases}{\label{Rheo}}
\ds
\sigma_{1\ell}(\varepsilon_{1\ell})=s(\varepsilon_{1\ell},\,K_{1\ell},\,{\bf p}),\label{Rheo1}\\
[6pt]
\ds
\sigma_{2\ell}(\varepsilon_{2\ell})=s(\varepsilon_{2\ell},\,K_{2\ell},\,{\bf 0}),\label{Rheo2}\\
\ds
\sigma_{3\ell}(\varepsilon_{2\ell})=\eta_\ell\frac{\textstyle \partial \varepsilon_{2\ell}}{\textstyle \partial t}\label{Rheo3}.
\end{subnumcases}
To determine the parameters $K_{1\ell}$, $K_{2\ell}$ and $\eta_\ell$, the relaxation times $\tau_{\sigma\ell}$, $\tau_{\varepsilon\ell}$ and the relaxed modulus $E_R$ are introduced:
\begin{equation}
\tau_{\sigma\ell}=\frac{\eta_\ell}{K_{1\ell}+K_{2\ell}},\hspace{0.5cm}
\tau_{\varepsilon\ell}=\frac{\eta_\ell}{K_{2\ell}},\hspace{0.5cm}
\frac{E_R}{N}=\frac{K_{1\ell}\,K_{2\ell}}{K_{1\ell}+K_{2\ell}}.
\label{TauEta}
\end{equation}
On the one hand, a procedure is given in \ref{SecViscoCoeff} to compute the relaxation times in terms of the quality factor $Q$. On the other hand, $E_R$ is related to the unrelaxed Young's modulus $E$ (\ref{Young}) and to the relaxation times previsously determined (see \cite{Carcione07}):
\begin{equation}
E_R=\frac{N}{\displaystyle \sum_{\ell=1}^N\frac{\tau_{\varepsilon\ell}}{\tau_{\sigma\ell}}}\,E.
\label{ErVsE}
\end{equation}
Once $\tau_{\sigma\ell}$, $\tau_{\varepsilon\ell}$ and $E_R$ are determined, inverting (\ref{TauEta}) provides the values of the viscoelastic model in terms of relaxed modulus and relaxation times ($\ell=1,\cdots,N$):
\begin{equation}
K_{1\ell}=\frac{\tau_{\varepsilon\ell}}{\tau_{\sigma\ell}}\,\frac{E_R}{N},\hspace{0.5cm}
K_{2\ell}=\frac{\tau_{\varepsilon\ell}}{\tau_{\varepsilon\ell}-\tau_{\sigma\ell}}\,\frac{E_R}{N},\hspace{0.5cm}
\eta_\ell=\frac{\tau^2_{\varepsilon\ell}}{\tau_{\varepsilon\ell}-\tau_{\sigma\ell}}\,\frac{E_R}{N}.
\label{E1E2Eta}
\end{equation}
From (\ref{ErVsE}) and (\ref{E1E2Eta}), it follows that the viscoelastic parameters depend indirectly on the Young's modulus $E$, and thus depend on $g$. In other words, the proposed model of viscoelasticity evolves with the concentration of defects and thus with the applied stress.

In the inviscid case, the stress-strain relation deduced from (\ref{MailleSigma})-(\ref{Rheo}) makes it possible to recover the nonlinear elasticity (\ref{ElastoNL}), whatever the number $N$ of relaxation mechanisms:
\begin{equation}
\sigma=s(\varepsilon,E,{\bf p}).
\label{Consistance}
\end{equation}
This property is proven in \ref{ProofHyp}.


\section{Mathematical modeling}\label{SecMath}

In this section, the basic components describing wave motion in damaged media are put together and analysed. Section \ref{SecMathEdp} collects the various mechanisms (nonlinear elastodynamics, slow dynamics, hysteresis) into a single system of first-order equations. Two important properties of this system are addressed in section \ref{SecMathProp}: hyperbolicity (finite sound velocity) and decrease in energy.

\subsection{First-order system}\label{SecMathEdp}

The conservation of momentum (\ref{ElastoLin1}) writes
\begin{equation}
\frac{\partial v}{\partial t}=\frac{1}{\rho}\frac{\partial \sigma}{\partial x}+\gamma,
\label{FO1}
\end{equation}
where $\gamma$ is a forcing term, and $\sigma$ is given by (\ref{MailleSigma}). The hypothesis of small deformations (\ref{ElastoLin2}) gives
\begin{equation}
\frac{\partial \varepsilon}{\partial t}=\frac{\partial v}{\partial x}.
\label{FO2}
\end{equation}
Lastly, manipulations on (\ref{MailleSigma}), (\ref{MailleEps}) and (\ref{Rheo3}) yield
\begin{equation}
\frac{\partial \varepsilon_{1\ell}}{\partial t}=\frac{\partial v}{\partial x}+\frac{\sigma_{2\ell}(\varepsilon-\varepsilon_{1\ell})-\sigma_{1\ell}(\varepsilon_{1\ell})}{\eta_\ell},\hspace{1cm}\ell=1,\cdots,N.
\label{FO3}
\end{equation}
In (\ref{FO3}), $\varepsilon_{1\ell}$ takes the place of the memory variables proposed in \cite{Lombard11} and is better suited to nonlinear elasticity. Putting together (\ref{FO1})-(\ref{FO3}) and the relaxation equation (\ref{Kinetic}) leads to the first-order system of $N+3$ evolution equations 
\begin{subnumcases}{\label{EDP}}
\frac{\partial v}{\partial t}-\frac{1}{\rho}\frac{\partial \sigma}{\partial x}=\gamma,\label{EDP1}\\
[4pt]
\frac{\partial \varepsilon}{\partial t}-\frac{\partial v}{\partial x}=0,\label{EDP2}\\
[4pt]
\frac{\partial \varepsilon_{1\ell}}{\partial t}-\frac{\partial v}{\partial x}=\frac{\sigma_{2\ell}(\varepsilon-\varepsilon_{1\ell})-\sigma_{1\ell}(\varepsilon_{1\ell})}{\eta_\ell}, \hspace{1cm}\ell=1,\cdots,\,N,\label{EDP3}\\
[4pt]
\frac{\textstyle dg}{\textstyle dt}=-\left(f_r\,H(g-g_\sigma)+f_d\,H(g_\sigma-g)\right)\,(g-g_\sigma).\label{EDP4}
\end{subnumcases}
To close the system (\ref{EDP}), the following equations are recalled:
\begin{itemize}
\item The total stress $\sigma$ in (\ref{EDP1}) depends on $\varepsilon_{1\ell}$ via (\ref{MailleSigma}), (\ref{Rheo1}), and a nonlinear law (\ref{ElastoNL}):
\begin{equation}
\sigma=\sum_{\ell=1}^N s(\varepsilon_{1\ell},\,K_{1\ell},\,{\bf p}).
\label{EDPsigma}
\end{equation}
\item The stress components $\sigma_{1\ell}$ and $\sigma_{2\ell}$ in (\ref{EDP3}) depend on the stifnesses $K_{1\ell}$ and $K_{2\ell}$ (\ref{Rheo1}) and (\ref{Rheo2}). The latter, as well as the viscosity coefficients $\eta_\ell$, depend on the Young modulus $E$ via (\ref{ErVsE})-(\ref{E1E2Eta}), and thus on $g$:
\begin{equation}
E=\left(1-\frac{\textstyle g}{\textstyle g_{cr}}\right)\,E^+.
\label{EDPYoung}
\end{equation}
\item The equilibrium value of the defect concentration $g_\sigma$ in (\ref{EDP4}) satisfies (\ref{GsigFLP}) and (\ref{Sig0}):
\begin{equation}
g_\sigma=\frac{\textstyle g_{cr}}{\textstyle 2}\left(1+\tanh\left(\frac{\textstyle \sigma-\sigma_c}{\textstyle \overline{\sigma}}\right)\right).
\label{EDPgsig}
\end{equation}
\end{itemize}
The system (\ref{EDP}), together with equations (\ref{EDPsigma})-(\ref{EDPgsig}), generalizes the standard equations of linear elastodynamics (\ref{ElastoLin}). It accounts for softening / recovering of Young's modulus, nonlinearity and viscoelasticity.

For the sake of clarity, the vector of $N+3$ variables is introduced
\begin{equation}
{\bf U}=\left(v,\,\varepsilon,\,\varepsilon_{11},\cdots,\,\varepsilon_{1N},\,g\right)^T.
\label{VecU}
\end{equation}
Then the system (\ref{EDP}) can be put in the form
\begin{equation}
\frac{\partial}{\partial t}{\bf U}+\frac{\partial}{\partial x}{\bf F}({\bf U})={\bf R}({\bf U})+{\bf \Gamma}.
\label{SysCons}
\end{equation}
The flux function ${\bf F}$, the relaxation term ${\bf R}$, and the forcing ${\bf \Gamma}$ are
\begin{equation}
\begin{array}{l}
\displaystyle
{\bf F}({\bf U})=\left(-\frac{\sigma}{\rho},\,-v,\,-v,\cdots,\,-v,0\right)^T,\\
[10pt]
\displaystyle
{\bf R}({\bf U})=\left(0,\,0,\,\Delta_1,\cdots,\,\Delta_N,-\left(f_r\,H(g-g_\sigma)+f_d\,H(g_\sigma-g)\right)\,(g-g_\sigma)
\right)^T,\\
[10pt]
\displaystyle
{\bf \Gamma}=\left(\gamma,\,0,\cdots,\,0,\,0\right)^T,
\label{FluxFunc}
\end{array}
\end{equation}
where
\begin{equation}
\Delta_\ell=\frac{\sigma_{2\ell}(\varepsilon-\varepsilon_{1\ell})-\sigma_{1\ell}(\varepsilon_{1\ell})}{\eta_\ell}.
\label{DeltaL}
\end{equation}
To conclude, let us consider the limit-case where the viscoelastic attenuation is neglected. In this case, equation (\ref{Consistance}) states that the stress-strain relations degenerate rigorously towards pure nonlinear elasticity, whatever $ N$. 


\subsection{Properties}\label{SecMathProp}

Hyperbolicity is a crucial issue in wave problems - physically, mathematically, and numerically. It amounts to saying that there exists a real and finite sound velocity $c$. This property was analysed in \cite{Ndanou14} for a particular nonlinear stress-strain relation in 3D. In 1D, it reduces to a simpler case detailed as follows.
Let us define the sound speed $c$ by
\begin{equation}
c^2=\sum_{\ell=1}^Nc_\ell^2=\frac{1}{\rho}\sum_{\ell=1}^N\frac{\partial \sigma_{1\ell}}{\partial \varepsilon_{1\ell}}.
\label{C2}
\end{equation}
The system (\ref{SysCons}) is hyperbolic if and only if $c^2>0$ in (\ref{C2}). 
The proof, as well as sufficient conditions on the strain to ensure hyperbolicity, is given in \ref{ProofHyp}. From (\ref{C2}), the local elastic modulus $M$ can be deduced:
\begin{equation}
M=\rho\,c^2=\sum_{\ell=1}^N\frac{\partial \sigma_{1\ell}}{\partial \varepsilon_{1\ell}}.
\label{ModuleC}
\end{equation}
Note that the Stokes viscoelastic model used in \cite{Vakhnenko05} introduces a term $\frac{\partial^2 v}{\partial x^2}$ in the right-hand side of (\ref{EDP3}). This Laplacian term destroys the hyperbolic character of the system (\ref{SysCons}). The viscoelastic model used here has therefore better mathematical properties.


Now let us examine the spectrum of the relaxation function in (\ref{SysCons}). 
Let us consider linear stress-strain relations. The parameters $K_{1\ell}$, $K_{2\ell}$ and $\eta_{\ell}$ are "freezed" in (\ref{ErVsE})-(\ref{E1E2Eta}), so that they do not depend on $g$ via $E$ (\ref{Young}). Then, the eigenvalues of the Jacobian matrix ${\bf J}=\frac{\partial {\bf R}}{\partial {\bf U}}$ are 
\begin{equation}
\mathrm{Sp}({\bf J})=\left\{0^2,\,-f_\xi,\,-\frac{K_{1\ell}+K_{2\ell}}{\eta_\ell}\right\}=\left\{0^2,\,-f_\xi,\,-\frac{1}{\tau_{\sigma\ell}}\right\},\hspace{0.5cm}\ell=1,\,\cdots,\,N,
\label{SpectreJ}
\end{equation}
(see (\ref{TauEta})), with $f_\xi=f_r$ if $g>g_\sigma$, $f_\xi=f_d$ if $g<g_\sigma$, $f_\xi=0$ else.
The proof is detailed in \ref{ProofRelax}. Two observations can be made:
\begin{itemize}
\item ${\bf J}$ is definite-negative if the relaxation frequencies $\tau_{\sigma\ell}$ are positive. The latter parameters are deduced from an optimization process based on the quality factor (\ref{SecViscoCoeff}). To ensure the energy decrease, it is therefore crucial to perform nonlinear optimization with constraint of positivity.
\item The optimization procedure detailed in \ref{SecViscoCoeff} is performed on the frequency range $[f_{\min},\,f_{\max}]$ surrounding the excitation frequency $f_c$. These frequencies satisfy
\begin{equation}
f_{\min}\approx \frac{1}{\max \tau_{\sigma\ell}}<f_c<f_{\max}\approx \frac{1}{\min \tau_{\sigma\ell}}.
\label{SpectreRange}
\end{equation}
In (\ref{SpectreRange}), $\approx$ are replaced by equalities if a linear optimisation is used \cite{Lombard11}. From (\ref{TimeScales}), it follows the spectral radius of ${\bf J}$ 
\begin{equation}
\varrho({\bf J})=\frac{1}{\min \tau_{\sigma\ell}} \gg f_\xi,
\label{Rspectre}
\end{equation}
so that the system (\ref{SysCons}) is stiff.
\end{itemize} 


\section{Numerical modeling}\label{SecNum}

In this section, a numerical strategy is proposed to integrate the first-order equations (\ref{SysCons}). For the sake of efficiency, a splitting approach is followed in section \ref{SecNumSplit}. The original equations are splitted into two parts, solved successively: a propagative part (section \ref{SecNumHyp}) and a relaxation part (section \ref{SecNumRelax}).

\subsection{Splitting}\label{SecNumSplit}

To integrate (\ref{SysCons}), a uniform spatial mesh $\Delta x$ and a variable time step $\Delta t^{(n)}\equiv \Delta t$ are introduced. An approximation ${\bf U}_i^n$ of the exact solution ${\bf U}(x_i=i\,\Delta x,\,t_n=t_{n-1}+\Delta t)$ is sought. A first strategy is to discretize explicitly the non-homogeneous system (\ref{SysCons}). But numerical stability implies a bound of the form
\begin{equation}
\Delta t\leq \min\left(\frac{\Delta x}{c_{\max}},\,\frac{2}{\varrho({\bf J})}\right),
\label{CFLdirect}
\end{equation}
where $c_{\max}=\max c_i^n$ is the maximal sound velocity at time $t_n$, and $\varrho({\bf J})$ is the spectral radius of the Jacobian of the relaxation term. As deduced from (\ref{Rspectre}), the second bound in (\ref{CFLdirect}) is penalizing compared with the standard CFL condition $\Delta t\leq \Delta x / c_{\max}$.

Here we follow another strategy: equation (\ref{SysCons}) is split into a hyperbolic step
\begin{equation}
\frac{\partial}{\partial t}{\bf U}+\frac{\partial}{\partial x}{\bf F}({\bf U})={\bf 0}
\label{SplittingHyp}
\end{equation}
and a relaxation step
\begin{equation}
\frac{\partial}{\partial t}{\bf U}={\bf R}({\bf U})+{\bf \Gamma}.
\label{SplittingRelax}
\end{equation}
The discrete operators associated with the discretization of (\ref{SplittingHyp}) and (\ref{SplittingRelax}) are denoted ${\bf H}_h$ and ${\bf H}_r$, respectively. The second-order Strang splitting is used, solving successively (\ref{SplittingHyp}) and (\ref{SplittingRelax}) with adequate time increments:
\begin{subnumcases}{\label{Splitting}}
\ds
{\bf U}_i^{(1)}={\bf H}_r\left(\frac{\Delta t}{2}\right)\,{\bf U}_i^{n},\label{Splitting1}\\
[6pt]
\ds
{\bf U}_i^{(2)}={\bf H}_h\left(\Delta t\right)\,{\bf U}_i^{(1)},\label{Splitting2}\\
[6pt]
\ds
{\bf U}_i^{n+1}={\bf H}_r\left(\frac{\Delta t}{2}\right)\,{\bf U}_i^{(2)}.\label{Splitting3}
\end{subnumcases}
Provided that ${\bf H}_h$ and ${\bf H}_r$ are second-order accurate and stable operators, the time-marching (\ref{Splitting}) gives a second-order accurate approximation of the original equation (\ref{SysCons}) \cite{Leveque02}.


\subsection{Hyperbolic step}\label{SecNumHyp}

The homogeneous equation (\ref{SplittingHyp}) is solved by a conservative scheme for hyperbolic systems \cite{Leveque02}
\begin{equation}
{\bf U}_i^{n+1}={\bf U}_i^n-\frac{\textstyle \Delta t}{\textstyle \Delta x}\left({\bf F}_{i+1/2}-{\bf F}_{i-1/2}\right).
\label{Conservatif}
\end{equation}
Many sophisticated schemes can be used for this purpose \cite{LeVeque03}. For the sake of simplicity and robustness, the Godunov scheme is used here. The numerical flux function ${\bf F}_{i+1/2}$ is computed using the Rusanov method \cite{Toro99}
\begin{equation}
{\bf F}_{i+1/2}=\frac{\textstyle 1}{\textstyle 2}\left({\bf F}({\bf U}_{i+1}^n)+{\bf F}({\bf U}_{i}^n)-\lambda_{i+1/2}^n({\bf U}_{i+1}^n-{\bf U}_i^n)\right),
\label{Rusanov}
\end{equation}
where ${\bf F}$ is the flux function (\ref{FluxFunc}), and the diffusion parameter $\lambda_{i+1/2}^n$ is given by the Davis approximation \cite{Davis88}
\begin{equation}
\lambda_{i+1/2}^n=\max\left(c_i^n,c_{i+1}^n\right).
\label{Davis}
\end{equation}
The Godunov scheme is first-order accurate and stable under the usual Courant-Friedrichs-Lewy (CFL) condition 
\begin{equation}
\Delta t= \frac{\alpha\,\Delta x}{c_{\max}},\mbox{  with } \alpha \leq 1.
\label{CFL}
\end{equation}


\subsection{Relaxation step}\label{SecNumRelax}

Let us denote $\overline{{\bf U}}=(\varepsilon,\,\varepsilon_{11},\cdots,\,\varepsilon_{1N})$ and $\overline{{\bf R}}$ the restriction of ${\bf R}({\bf U})$ to the strain components (\ref{FluxFunc})-(\ref{DeltaL}). The ordinary differential equation (\ref{SplittingRelax}) can then be written
\begin{subnumcases}{\label{ODE}}
\frac{\partial v}{\partial t}=\gamma,\label{ODE1}\\
[4pt]
\frac{\partial}{\partial t}\overline{{\bf U}}=\overline{{\bf R}}(\overline{{\bf U}}),\label{ODE2}\\
[4pt]
\frac{\textstyle dg}{\textstyle dt}=-\left(f_r\,H(g-g_\sigma)+f_d\,H(g_\sigma-g)\right)\,(g-g_\sigma),\label{ODE3}
\end{subnumcases}
The viscoelastic parameters in the relaxation function $\overline{{\bf R}}$ depend implicitly on $g$ (see section \ref{SecPhysElasto}), which complicates the resolution of (\ref{ODE1}). However, one can take advantage of the scaling (\ref{TimeScales}). Indeed, $\varepsilon$ and $\varepsilon_{1\ell}$ evolve much faster than $g$, so that the viscoelastic parameters $K_{1\ell}$, $K_{2\ell}$, $\eta_\ell$ are almost constant on a time step. Consequently, they are "freezed" and the three equations in (\ref{ODE}) can be solved separately.

The half-time step in the relaxation steps (\ref{Splitting1})-(\ref{Splitting3}) is denoted by $\tau=\frac{\Delta t}{2}$. One details the time-stepping from $t_n$ to the first intermediate step (\ref{Splitting1}); adaptation to the third intermediate step (\ref{Splitting3}) is straightforward. 

The first equation (\ref{ODE1}) is integrated using the Euler method:
\begin{equation}
v_i^{n+1}=v_i^{(1)}+\Delta t\,\gamma(i,\,t_n).
\label{EulerODE2}
\end{equation}
To integrate the second equation (\ref{ODE2}), a first-order Taylor expansion of $\overline{{\bf R}}(\overline{{\bf U}})$ is performed 
\begin{equation}
\frac{\partial}{\partial t}\overline{{\bf U}}\approx\overline{{\bf R}}(\overline{{\bf 0}})+\frac{\partial \overline{{\bf R}}}{\partial \overline{{\bf U}}}({\bf 0})\,\overline{{\bf U}}={\bf \overline{J}}\,\overline{{\bf U}},
\label{SemiLinear}
\end{equation}
where ${\bf \overline{J}}$ is the Jacobian matrix (\ref{MatJ}); the nullity of stress at zero strain has been used (\ref{Rheo1}). Then (\ref{SemiLinear}) is solved exactly, leading to the relaxation operator 
\begin{equation}
\overline{{\bf U}}^{(1)}_i=e^{{\bf \overline{J}}\,\tau}\,\overline{{\bf U}}^{n}_i
\label{ExpoS1}
\end{equation}
with the matrix exponential   
\begin{equation}
e^{{\bf \overline{J}}\,\tau}=
\left(
\begin{array}{cccc}
1   & 0 & \cdots & 0\\
[8pt]
\ds \frac{E_{21}}{E_{11}+E_{21}}\left(1-e^{-\frac{E_{11}+E_{21}}{\eta_1}\,\tau}\right) & \ds e^{-\frac{E_{11}+E_{21}}{\eta_1}\,\tau} &  &  \\
\vdots & & \ddots & \\
\ds \frac{E_{2N}}{E_{1N}+E_{2N}}\left(1-e^{-\frac{E_{1N}+E_{2N}}{\eta_N}\,\tau}\right) & & & e^{-\frac{E_{1N}+E_{2N}}{\eta_N}\,\tau}
\end{array}
\right).
\label{ExpoS2}
\end{equation}
Lastly, the third equation (\ref{ODE3}) is solved exactly. The grid value $g_{\sigma i}$ is evaluated thanks to (\ref{GsigFLP}). Setting 
\begin{equation}
f_\xi=
\left\{
\begin{array}{l}
f_r \mbox{  if } g_i^n\geq g_{\sigma i}^n,\\
[8pt]
f_d \mbox{  if } g_i^n< g_{\sigma i}^n,
\end{array}
\right.
\label{MuNu}
\end{equation}
leads to
\begin{equation}
g_i^{(1)}=g_{\sigma i}^n+\left(g_i^n-g_{\sigma i}^n\right)\,e^{-f_\xi\,\tau}.
\label{SolveODE3}
\end{equation}
The integrations (\ref{ExpoS1}), (\ref{EulerODE2}) and (\ref{SolveODE3}) are unconditionally stable. As a consequence, the splitting (\ref{Splitting}) is stable under the CFL condition (\ref{CFL}).


\subsection{Summary of the algorithm}\label{SecNumDiffu}

The numerical method can be divided in two parts:
\begin{enumerate}
\item initialisation
\begin{itemize}
\item bulk modulus $\rho$, Young's modulus $E=E_0=\rho\,c_\infty^2$;
\item soft-ratchet coefficients $g_{cr}=1$, $g=g_0$, $f_r$, $f_d$, $\overline{\sigma}$;
\item maximum Young's modulus $E^+$ (\ref{Young})
\item nonlinear coefficients (e.g. $\beta$ and $\delta$ in (\ref{Model3});
\item quality factor $Q$, frequency range of optimization $[f_{\min},\,f_{\max}]$, number of relaxation mechanisms $N$;
\item optimization of the viscoelastic coefficients (\ref{SecViscoCoeff});
\end{itemize}
\item time-marching $t_n \rightarrow t_{n+1}$, $x_i=i\,\Delta x$ ($n=0,\cdots,\,N_t$, $i=1,\cdots,\,N_x$)
\begin{itemize}
\item physical and numerical parameters
\begin{description}
\item[-] Young's modulus $E$ (\ref{Young}), viscoelastic parameters $E_R$ (\ref{ErVsE}), $K_{1\ell}$, $K_{2\ell}$ and $\eta_\ell$ (\ref{E1E2Eta});
\item[-] partial stresses $\sigma_{1\ell}$ (\ref{Rheo1}) and total stress $\sigma$ (\ref{MailleSigma});
\item[-] sound velocity $c$ (\ref{C2}) and (\ref{C2Modeles}), maximal velocity $c_{\max}$;
\item[-] time step $\Delta t$ (\ref{CFL});
\end{description}
\item relaxation step ${\bf H}_r$ (\ref{Splitting1})
\begin{description}
\item[-] strains (\ref{ExpoS1}) and (\ref{ExpoS2});
\item[-] velocity $v$ (\ref{EulerODE2});
\item[-] concentration of defects at equilibrium $g_\sigma$ (\ref{GsigFLP}) and out of equilibrium $g$ (\ref{SolveODE3});
\end{description}
\item hyperbolic step ${\bf H}_h$ (\ref{Splitting2})
\begin{description}
\item[-] coefficient $\lambda_{i+1/2}$ of Davis (\ref{Davis});
\item[-] computation of the flux ${\bf F}$ (\ref{FluxFunc}), e.g., by the Rusanov flux ${\bf F}_{i+1/2}$ (\ref{Rusanov});
\item[-] time-marching of the conservative scheme (\ref{Conservatif});
\end{description}
\item relaxation step ${\bf H}_r$ (\ref{Splitting3}).
\end{itemize}
\end{enumerate}


\section{Numerical experiments}\label{SecExp}

\subsection{Configuration}\label{SecExpConfig}

\begin{table}[htbp]
\begin{center}
\begin{small}
\begin{tabular}{|c|c|c|c|c|c|c|c|c|}
\hline
$\rho$ (kg/m$^{3}$) & $E_0$ (GPa) & $g_0$ & $f_r$ (Hz) & $f_d$ (Hz) & $\overline{\sigma}$ (GPa) & $\beta$ & $\delta$ & $Q$ \\
\hline
2054 & 2.21 & 0.1 & 25 & 250 & 0.1 & 40 & $3.5\,10^6$ & 20 \\
\hline
\end{tabular}
\end{small} 
\end{center}
\vspace{-0.5cm}
\caption{\label{TabParam} Physical parameters.}
\end{table}

The physical parameters are detailed in table \ref{TabParam}. Depending on the test, some of these parameters are modified. In the limit-case of linear elasticity, the sound velocity is $c=\sqrt{E/\rho}=3280$ m/s. The maximal CFL number is $\alpha=0.95$ in (\ref{CFL}). The mesh size is $\Delta x=4\,10^{-3}$ m. Depending on the test, two lengths of domain are considered. For each test, a receiver put at $x_r=0.2$ m stores the numerical solution at each time step.

The wave fields are excited by a punctual source at $x_s=10^{-2}$ m, with a central frequency $f_c=10$ kHz. Depending on the expression of the forcing $\gamma$ in (\ref{EDP3}), it is possible to deduce the magnitude of the maximal strain $\varepsilon_{\max}$ emitted by the source in the limit-case of linear elasticity (\ref{ElastoExact}):
\begin{equation}
\varepsilon_{\max}=\frac{1}{2\,c^2}\max{\cal G}(t).
\label{EpsMax}
\end{equation} 

The Landau model for nonlinear elasticity is used (\ref{Model3}). The coefficient $\beta$ is much smaller than $\delta$. The critical value of strain that ensures hyperbolicity (\ref{EpsC}) is $\varepsilon_c=3.08\,10^{-4}$. The viscoelastic effects are described by $N=4$ relaxation mechanisms. The relaxation times $\tau_{\sigma\ell}$ and $\tau_{\varepsilon\ell}$ (\ref{TauEta}) are computed by optimization on the frequency range $[f_{\min}=f_c/10,\,f_{\max}=f_c \times 10]$ (see \ref{SecViscoCoeff}); they are given in table \ref{TabRelax}. 

\begin{table}[h!]
\begin{center}
\begin{small}
\begin{tabular}{|c|c|c|c|c|}
\hline
& $\ell=1$ & $\ell=2$ & $\ell=3$ & $\ell=4$\\
\hline
$\tau_{\sigma\ell}$ (s) & $1.16 \,10^{-3}$ & $2.05\,10^{-4}$ & $4.49\,10^{-5}$ & $7.75\,10^{-6}$ \\
\hline
$\tau_{\varepsilon\ell}$ (s) & $1.53 \,10^{-3}$ & $2.49\,10^{-4}$ & $5.50\,10^{-5}$ & $1.06\,10^{-5}$ \\
\hline
\end{tabular}
\end{small} 
\end{center}
\vspace{-0.5cm}
\caption{\label{TabRelax} Relaxation times for a quality factor $Q=20$. Optimization with $N=4$ relaxation mechanisms on the frequency range [1 kHz, 100 kHz].}
\end{table}


\subsection{Test 1: nonlinear elastodynamics}\label{SecExpTest1}

\begin{figure}[h!]
\begin{center}
\begin{tabular}{cc}
$\varepsilon_{\max}=10^{-5}$ & $\varepsilon_{\max}=2.0\,10^{-4}$\\
\hspace{-0.9cm} \includegraphics[scale=0.35]{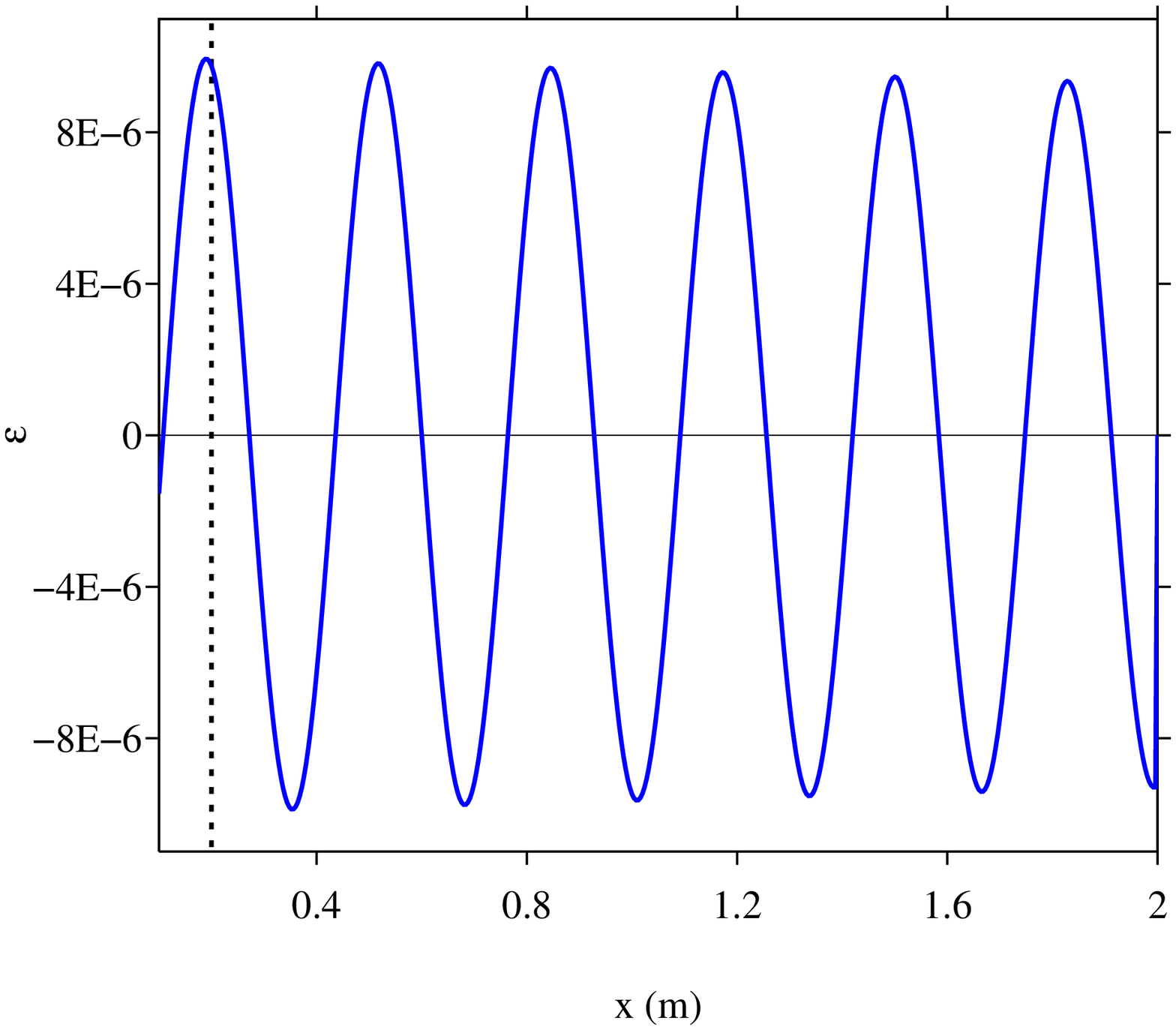} & 
\hspace{-0.9cm} \includegraphics[scale=0.35]{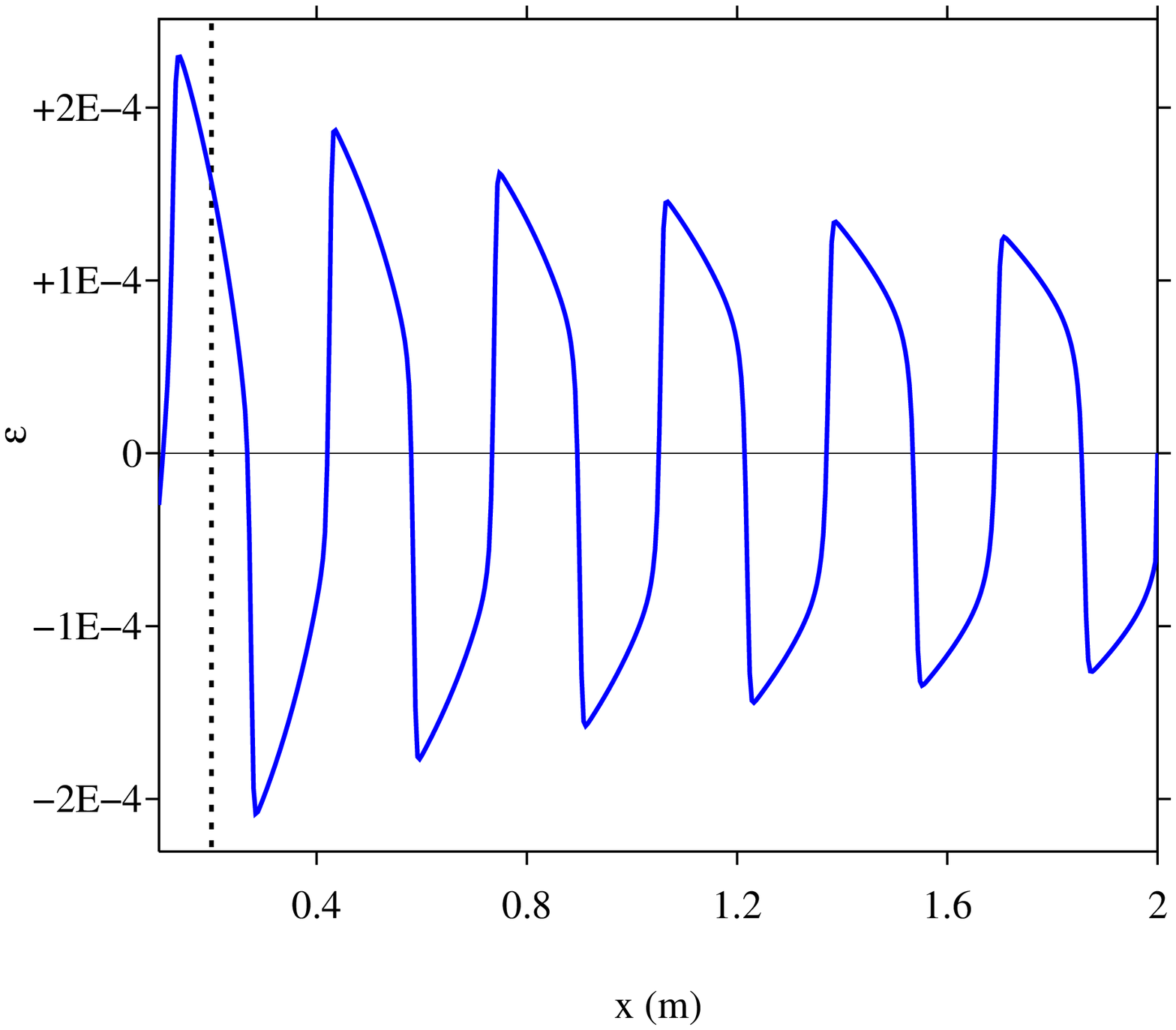} \\
\end{tabular}
\caption{\label{FigTest1X} test 1. Snapshot of the strain after 400 time steps, for two amplitudes of the excitation. The vertical dotted line denotes the location $x_r$ of the receiver.}
\end{center}
\end{figure}

In the first test, the viscoelasticity is neglected, and the activation / restoration of defects is annihilated: $f_r=f_d=0$ Hz. This test corresponds to the example 12 of \cite{Whitham74}. Our goal is to show typical features of wave propagation in purely nonlinear elastic media. The source is a monochromatic excitation:
\begin{equation}
{\cal G}(t)=A\,\sin(\omega_c t)\,H(t),
\label{SourceSinus}
\end{equation}
where $A$ is the magnitude of the forcing, and $\omega_c=2\,\pi\,f_c$. From (\ref{EpsMax}) and (\ref{SourceSinus}), it is possible to estimate the maximal strain $\varepsilon_{\max}$ emitted by the source in the linear elastic case. The domain of propagation is $L_x=2$ m long and is discretized onto 400 grid nodes. 

Figure \ref{FigTest1X} displays the spatial evolution of $\varepsilon$ after 400 time steps. For $\varepsilon_{\max}=10^{-5}$, almost no distorsion of the wave is seen. On the contrary, $\varepsilon_{\max}=2.0\,10^{-4}$ yields a high distorsion as the wave propagates. Shocks, as well as the attenuation due to the intersection of characteristic curves \cite{Leveque02}, are observed.

\begin{figure}[h!]
\begin{center}
\begin{tabular}{cc}
(a) & (b)\\
\hspace{-0.9cm} \includegraphics[scale=0.35]{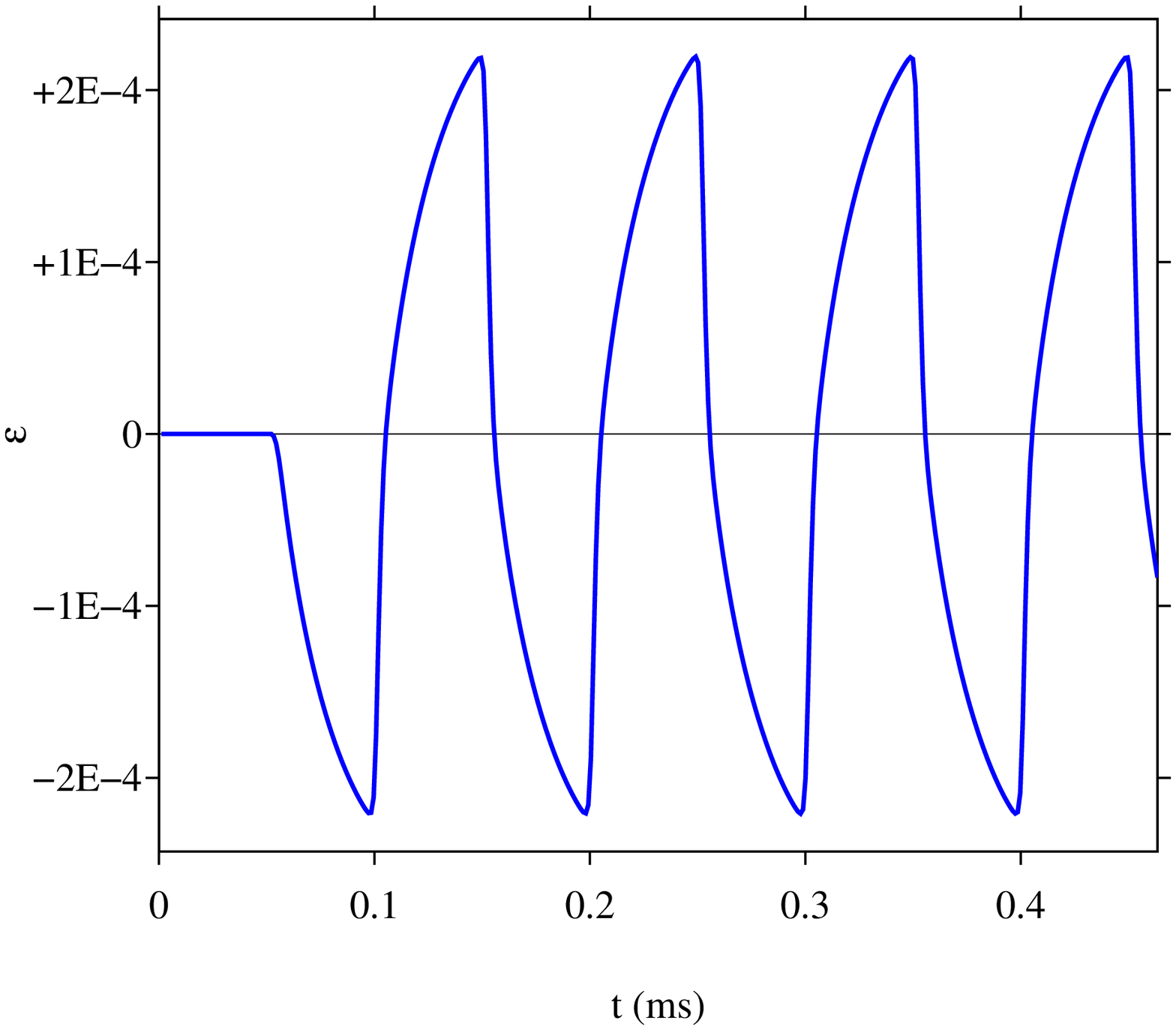} & 
\hspace{-0.9cm} \includegraphics[scale=0.35]{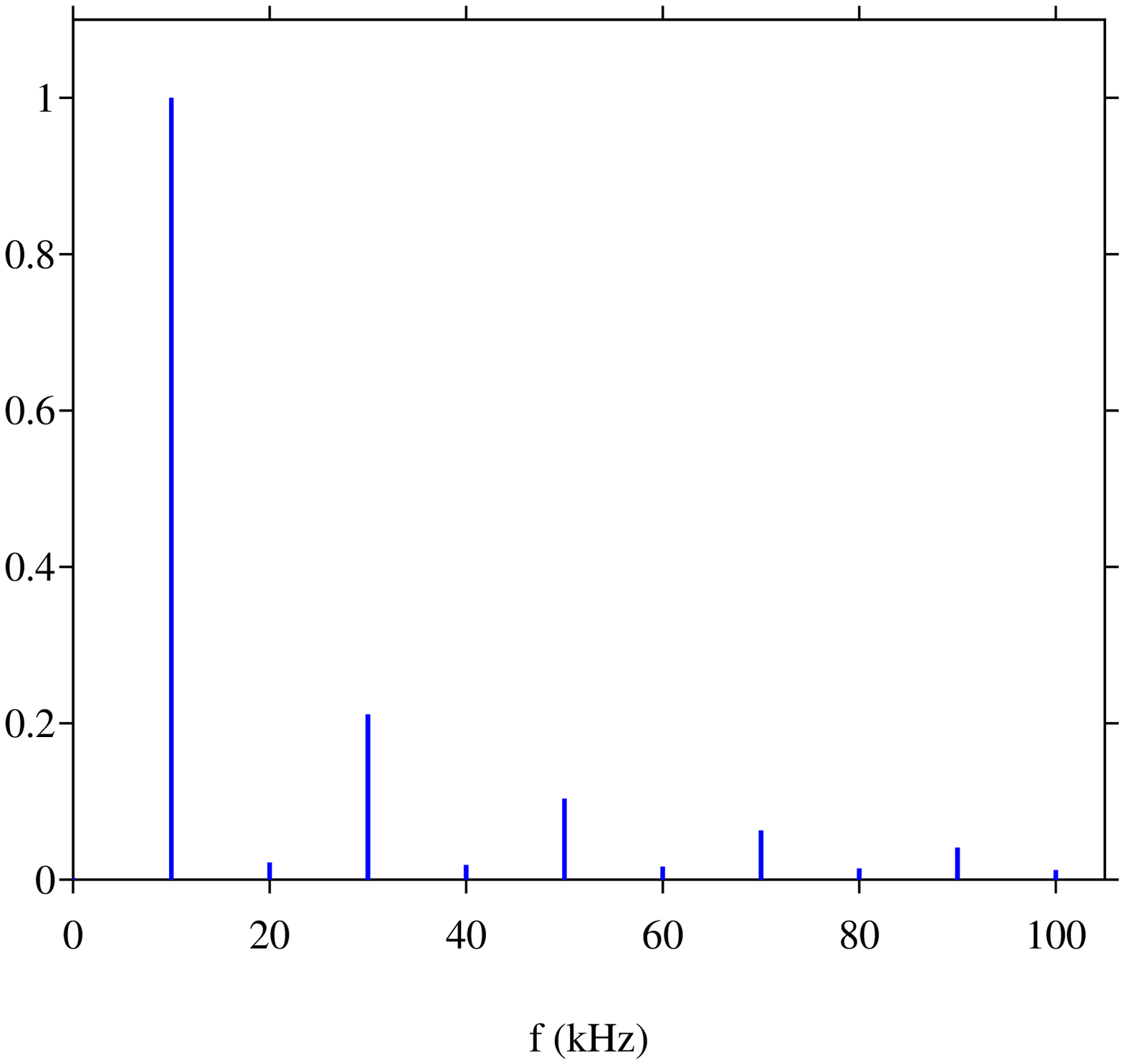} \\
\end{tabular}
\caption{\label{FigTest1T} test 1. Time history of the strain at the receiver at $x_r$ (a), normalized Fourier coefficients (b). The amplitude of the excitation is $\varepsilon_{\max}=2.0\,10^{-4}$.}
\end{center}
\end{figure}

Figure \ref{FigTest1T} displays the time evolution of the strain recorded at the receiver (vertical dotted line in Figure \ref{FigTest1X}) for $\varepsilon_{\max}=2.0\,10^{-4}$. The normalized amplitudes of the Fourier series decomposition show a typical feature of cubic nonlinear elasticity: the spectrum involves mainly odd harmonics \cite{Hamilton98}.


\subsection{Test 2: linear viscoelasticity}\label{SecExpTest2}

The goal of the second test is to validate the numerical modeling of attenuation. For this purpose, a linear stress-strain relation is chosen ($\beta=\delta=0$), and the activation / restoration of defects is still annihilated ($f_r=f_d=0$ Hz). Consequently, the system (\ref{EDP}) simplifies into
\begin{subnumcases}{\label{EDPVisco}}
\frac{\partial v}{\partial t}-\frac{1}{\rho}\frac{\partial \sigma}{\partial x}=\gamma,\label{EDPVisco1}\\
[4pt]
\frac{\partial \varepsilon}{\partial t}-\frac{\partial v}{\partial x}=0,\label{EDPVisco2}\\
[4pt]
\frac{\partial \varepsilon_{1\ell}}{\partial t}-\frac{\partial v}{\partial x}=\frac{K_{2\ell}}{\eta_\ell}(\varepsilon-\varepsilon_{1\ell})-\frac{K_{1\ell}}{\eta_\ell}\varepsilon_{1\ell}.\label{EDPVisco3}
\end{subnumcases}
The domain of propagation is $L_x=2$ m long and is discretized onto 400 grid nodes. The time evolution of the source is a truncated combination of sinusoids with $C^6$ smoothness:
\begin{equation}
{\cal G}(t) = 
\left\{
\begin{array}{l}
\ds
\sum_{m=1}^4 a_m \sin\,(b_m\,\omega_c\,t)  \mbox{  if }\;0\leq t\leq \frac{1}{f_c},\\
[6pt]
\ds
0 \qquad \mbox{otherwise},
\end{array}
\right. 
\label{JKPS_C6}
\end{equation}
with parameters $b_m=2^{m-1}$, $a_1=1$, $a_2=-21/32$, $a_3=63/768$ and $a_4=-1/512$. Five receivers are put at abscissae $x_r=0.5+0.3\,(j-1)$, with $j=1,\,\cdots 5$. 

\begin{figure}[h!]
\begin{center}
\begin{tabular}{cc}
(a) & (b)\\
\hspace{-0.9cm} \includegraphics[scale=0.35]{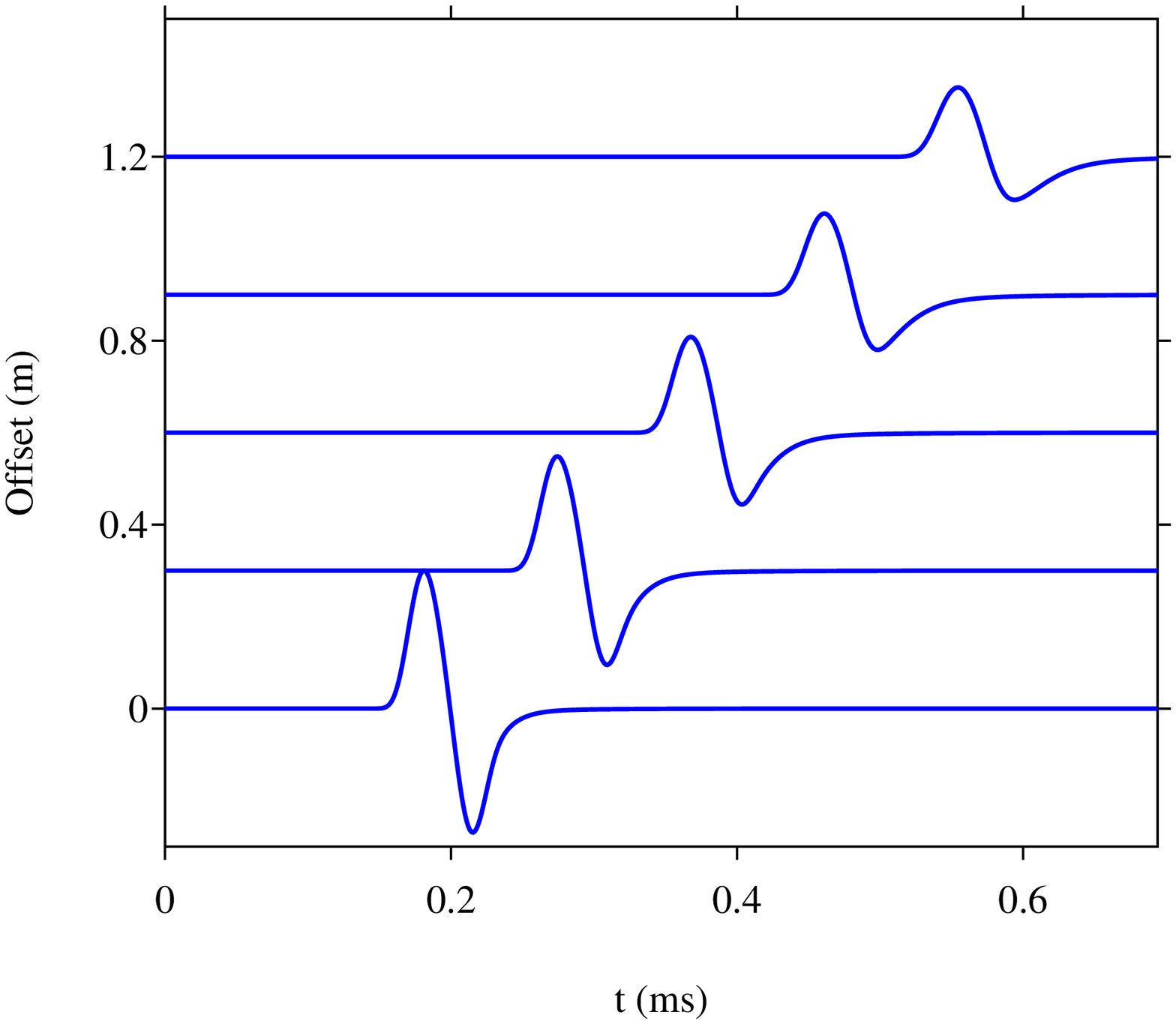} & 
\hspace{-0.9cm} \includegraphics[scale=0.35]{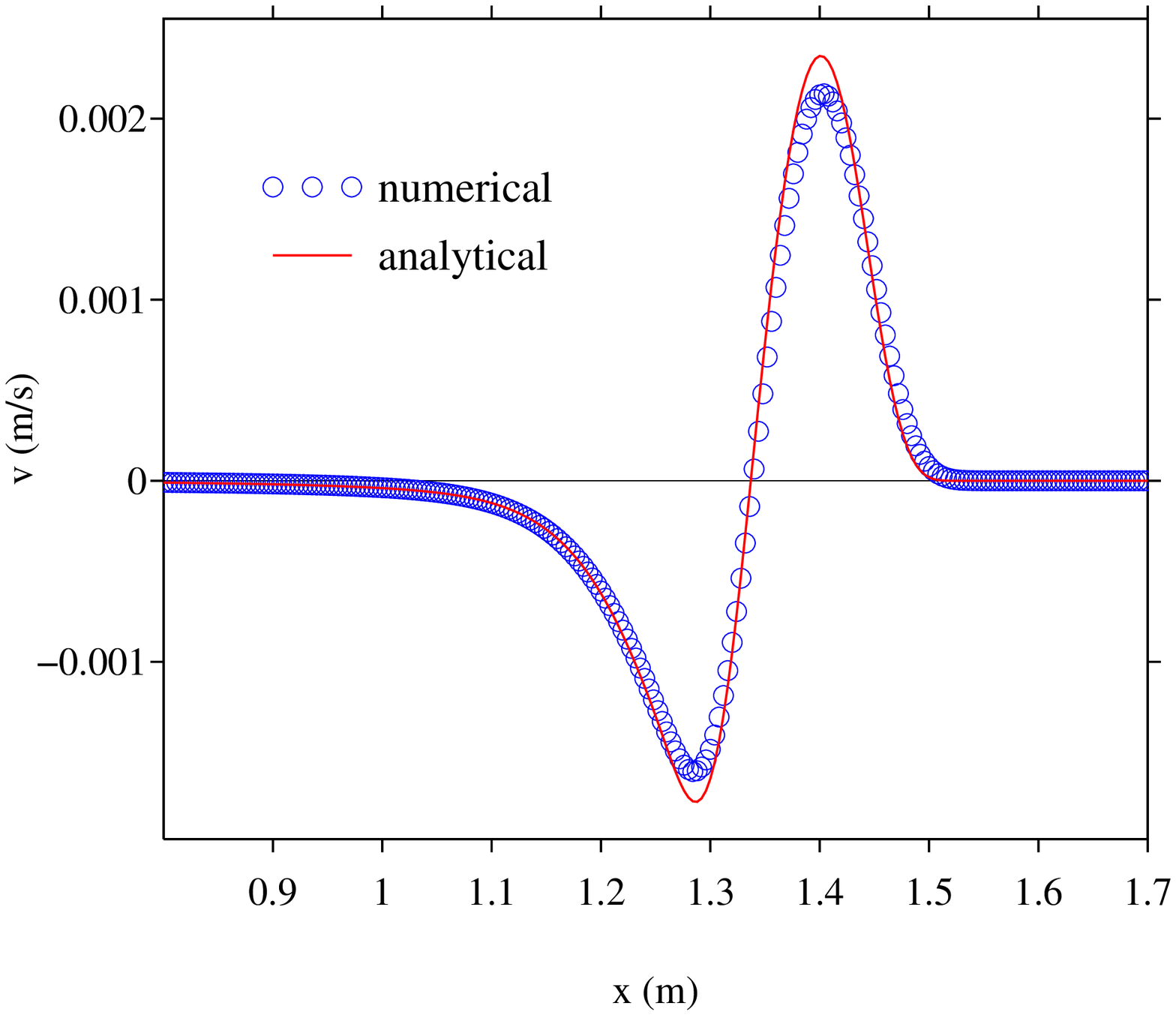} 
\end{tabular}
\caption{\label{FigTest2} test 2. Wave propagation in a viscoelastic medium. (a): time evolution of $v$ at a set of receivers; (b): snapshot of $v$ at $t=0.46$ ms, and comparison between the numerical and the semi-analytical solution.}
\end{center}
\end{figure}

Figure \ref{FigTest2}-(a) shows a seismogram of the velocity recorded at the receivers. Attenuation and dispersion of the waves is clearly observed. Figure \ref{FigTest2}-(b) compares the numerical solution with the semi-analytical solution after 400 time steps. The computation of the semi-analytical solution is described in \ref{SecSolExact}; it is numerically evaluated with $N_f=512$ Fourier modes, with a frequency step $\Delta f=200$ Hz. Good agreement is observed between numerical and exact values. The attenuation is slightly overestimated by the scheme, due to the numerical diffusion of the Godunov scheme. This numerical artifact can be fixed by choosing a higher-order scheme \cite{Toro99}.


\subsection{Test 3: softening / recovering}\label{SecExpTest3}

The goal of the third test is to illustrate the softening / recovering of the elastic modulus, and to validate the numerical modeling of this phenomenon. For this purpose, linear elasticity is assumed and the viscoelasticity is neglected ($\beta=\delta=0$, $Q=+\infty$). Even if a linear stress-strain relation is used, the evolution problem (\ref{EDP}) is nonlinear by virtue of (\ref{EDP4}), (\ref{EDPYoung}) and (\ref{EDPgsig}). Like in test 1, the source is monochromatic; but is is switched off after a time $t^*$:
\begin{equation}
{\cal G}(t)=A\,\sin(\omega_c t)\,\left(H(t)-H(t^*)\right).
\label{SourceSwitch}
\end{equation}
As long as the source is switched on ($0<t<t^*$), the equilibrium concentration of defects increases from the initial value $g_0$ up to $g^*=g(t^*)$. At the same time, the Young's modulus decreases from $E_0$ to $E^*$ via (\ref{Young}).

\begin{figure}[htbp]
\begin{center}
\begin{tabular}{cc}
(a) & (b)\\
\hspace{-0.9cm}
\includegraphics[scale=0.42]{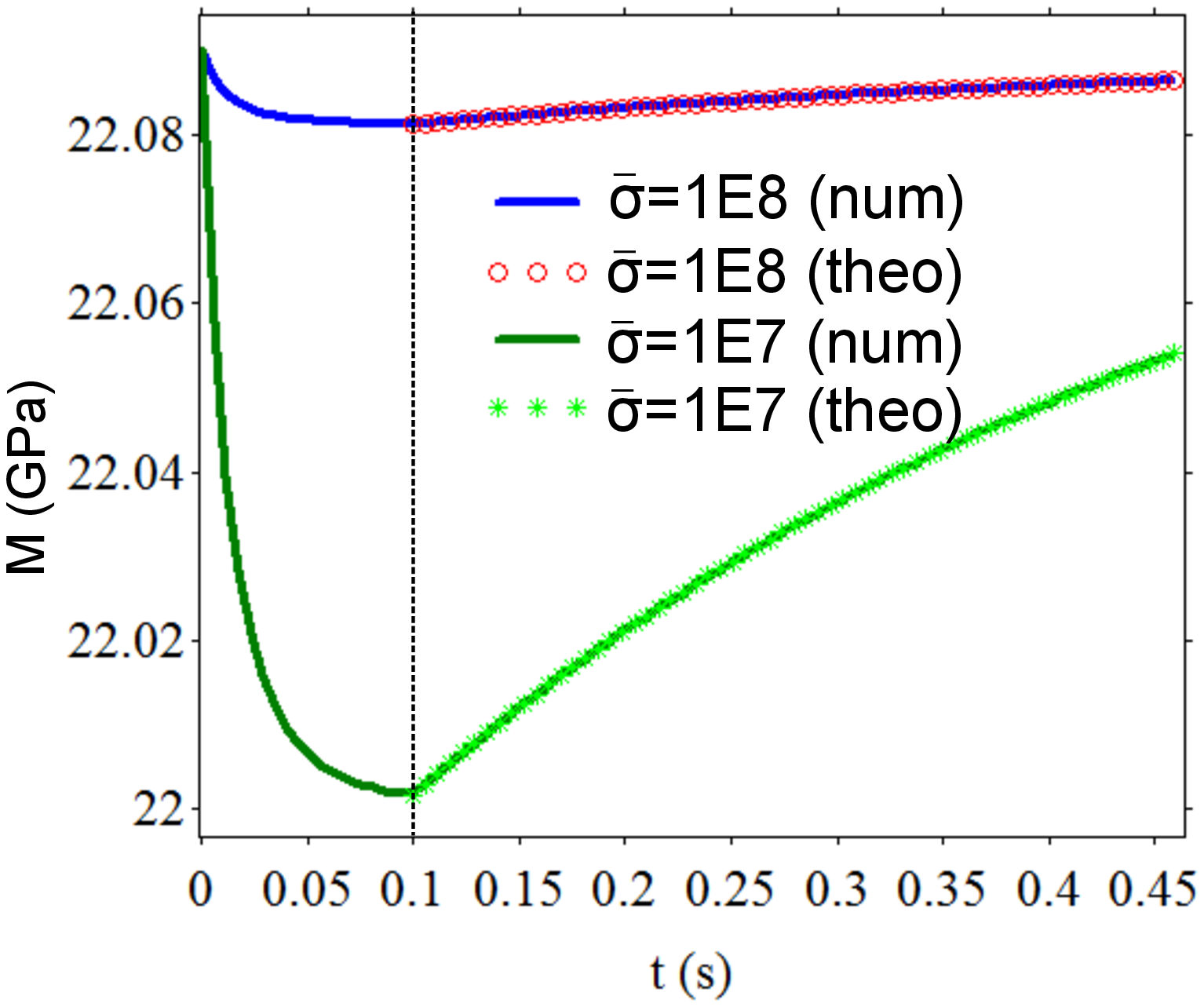} & 
\hspace{-0.3cm}
\includegraphics[scale=0.42]{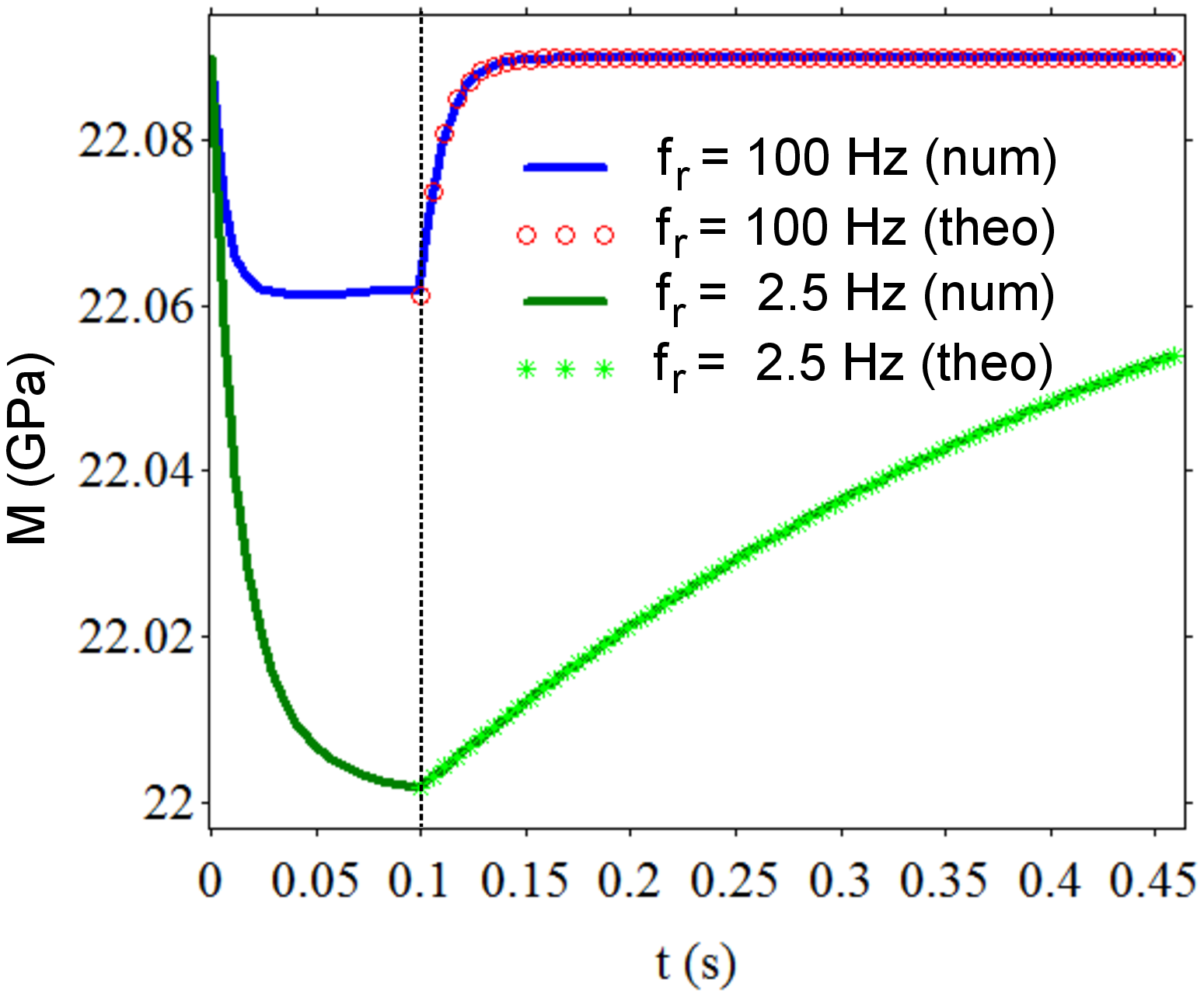} 
\end{tabular}
\caption{\label{FigTest3} test 3. Time evolution of the elastic modulus $M$ (\ref{ModuleC}) at $x_r$. (a): influence of the central stress $\overline{\sigma}=10^8$ Pa and $10^7$ Pa. (b): influence of the frequency of restoration $f_r=2.5$ Hz and 100 Hz. The vertical dotted line denotes the time $t^*$ where the source is switched off.}
\end{center}
\end{figure}

For $t>t^*$, the waves go out of the domain, and the elastodynamic fields vanish. From (\ref{EDPgsig}) and (\ref{Sig0}), $\sigma=0$ implies that the equilibrium concentration of defects becomes $g_\sigma=g_0$. As a consequence, the ordinary differential equation (ODE) (\ref{EDP4}) describing the evolution of defects simplifies into
\begin{equation}
\left\{
\begin{array}{l}
\ds
\frac{\textstyle dg}{\textstyle dt}=-f_r\,(g-g_0),\\
[8pt]
\ds
g(t^*)=g^*.
\end{array}
\right.
\label{ODEgstar}
\end{equation}
The solution of (\ref{ODEgstar}) is 
\begin{equation}
g(t)=g_0+\left(g^*-g_0\right)\,e^{-f_r (t-t^*)}.
\label{gstar}
\end{equation}
Equation (\ref{gstar}) is injected into (\ref{Young}), which gives the time evolution of the Young's modulus during the recovering process ($t\geq t^*$):
\begin{equation}
E(t)=E_0-\frac{1}{g_{cr}}\left(g^*-g_0\right)\,e^{-f_r (t-t^*)}\,E^+.
\label{Estar}
\end{equation}
The domain of propagation is $L_x=0.4$ m long and is discretized onto 100 grid nodes. The maximal strain is $\varepsilon_{\max}=10^{-5}$. Time integration is performed up to $t=460$ ms. Figure \ref{FigTest3} shows the time evolution of the elastic modulus $M\equiv E$ (\ref{ModuleC}); this equality occurs only because a linear stress-strain relation is assumed. The numerical values of $M$ are shown from the beginning of the simulation, whereas the exact values of $E$ (\ref{Estar}) are shown from $t^*$. For the sake of clarity, the values are shown only each 5000 time steps. Logically, the elastic modulus decreases as long as the source is switched on (softening), and then increases up to its initial value (recovering).

Figure \ref{FigTest3}-(a) illustrates the influence of the central stress on the evolution of $M$: $\overline{\sigma}=10^8$ Pa or $10^7$ Pa (the other parameters are those of table \ref{TabParam}). According to the Vakhnenko's expression (\ref{GsigVakh}), these values correspond to spherical defects of radius $2.13\,10^{-10}$ m and $4.59\,10^{-10}$ m, respectively. In both cases, equilibrium has been reached at $t^*$. The lower value of $\overline{\sigma}$ yields a greater variation of the elastic modulus. This property follows from (\ref{GsigFLP}): as $\overline{\sigma}$ decreases, the curve $g\rightarrow g_\sigma$ stiffens and tend towards a Heaviside step function. Consequently, greater values of $g_\sigma$ are obtained when $\overline{\sigma}$ is smaller. This implies a greater evolution of $g$ (\ref{Kinetic}), and hence of $E$ (\ref{Young}).

Figure \ref{FigTest3}-(b) illustrates the influence of the frequency of restoration on the evolution of $M$: $f_r=2.5$ Hz or 100 Hz (the other parameters are those of table \ref{TabParam}). The lowest value of $f_r$ yields a greater variation of the elastic modulus. This is a consequence of the competition between restoration (with frequency $f_r$) and destruction (with frequency $f_d$). When $f_r$ is too low compared with $f_d$, restoration has almost no time to occur during one period $T=1/f_c$, and destruction plays a preponderant role.


\subsection{Test 4: full model}\label{SecExpTest4}

\begin{figure}[h!]
\begin{center}
\begin{tabular}{cc}
(a) & (b)\\
\hspace{-0.9cm}
\includegraphics[scale=0.44]{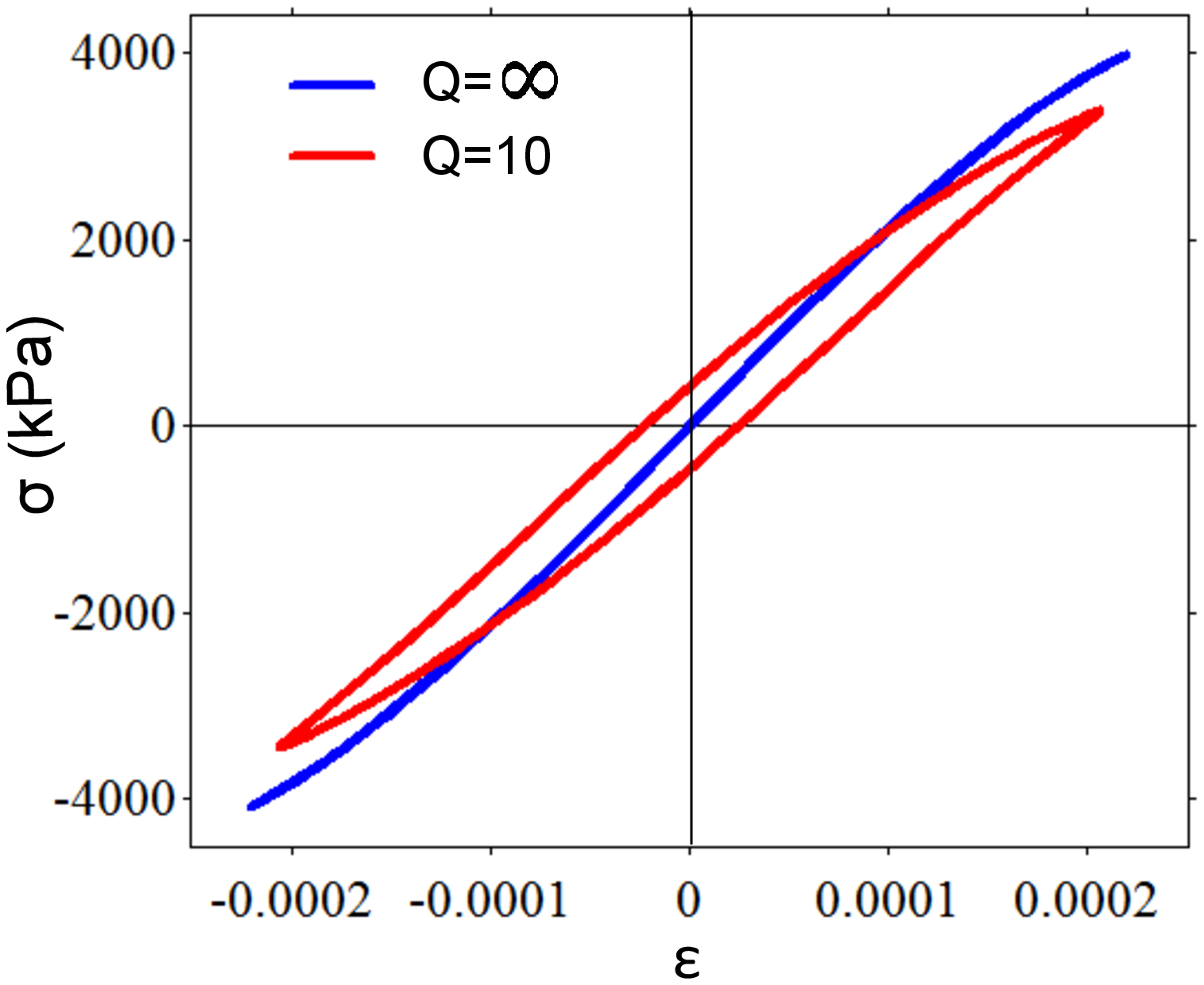} &
\hspace{-0.9cm}
\includegraphics[scale=0.35]{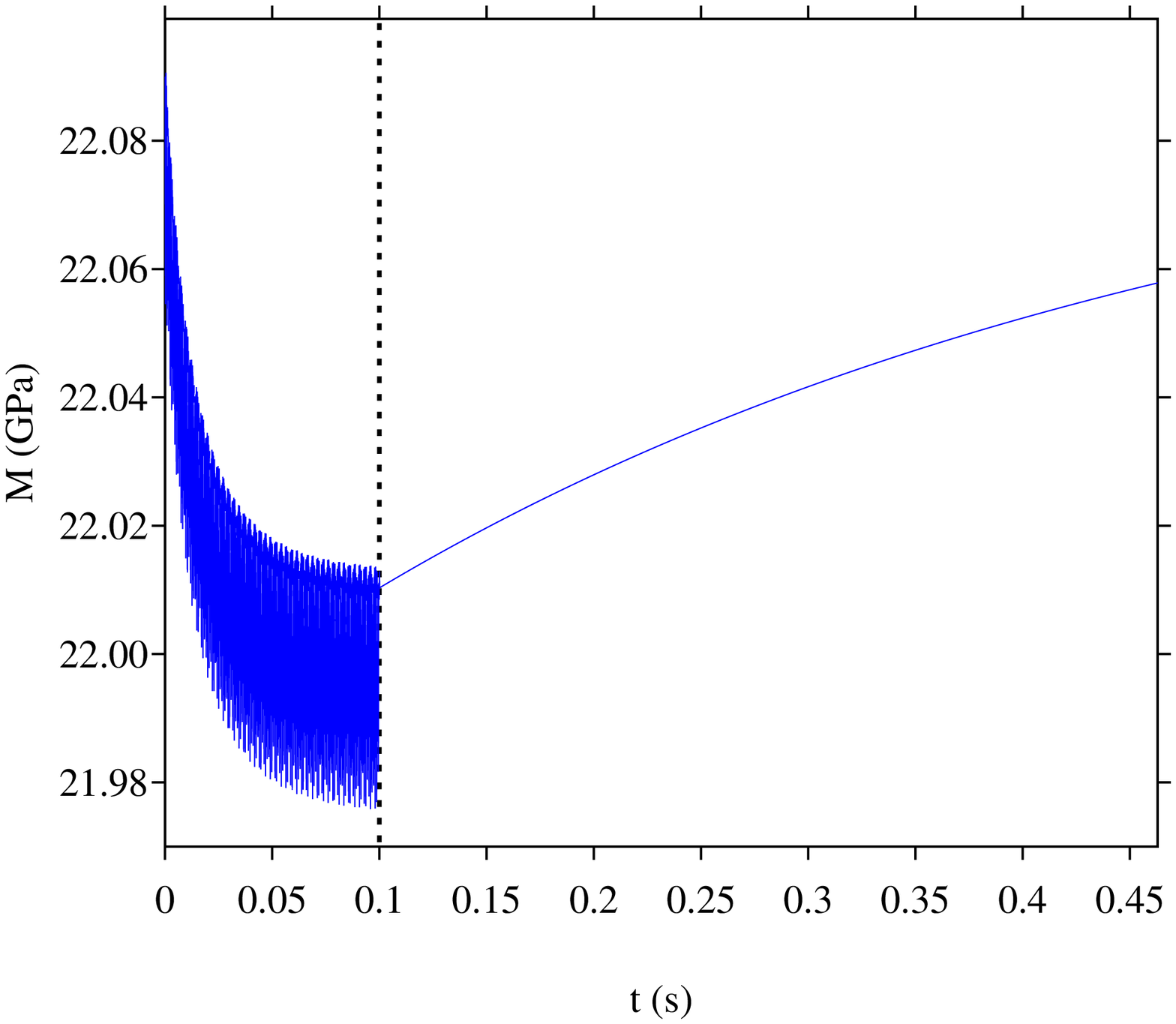} 
\end{tabular}
\caption{\label{FigSigvsEps} test 4. (a): stress-strain curves at $x_r$ for different quality factor $Q$ and a forcing amplitude $\varepsilon_{\max}=2.0\,10^{-4}$. (b): time evolution of the elastic modulus; the vertical dotted line denotes the time $t^*$ when the source is switched off.}
\end{center}
\end{figure}

The fourth and last test incorporates all the physical mechanisms of the model: nonlinear stress-strain law, viscoelasticity, activation / restoration of defects. The domain is $L_x=0.4$ m long and is discretized onto 100 grid nodes. The source is a monochromatic excitation (\ref{SourceSinus}). Time integration is performed during $5\,10^4$ time steps. The fields are recorded at $x_r$. 

Figure \ref{FigSigvsEps}-(a) illustrates the influence of viscoelasticity on the stress-strain law. When viscous effects are neglected ($Q=+\infty$, where $Q$ is the quality factor), the behavior induced by the Landau law (\ref{Model3}) is observed. Moreover, the scaling (\ref{TimeScales}) induces that the evolution of defects on one cycle is insufficient to provide a measurable hysteretic effect. On the contrary, hysteresis is obtained when viscoelasticity is accounted for ($Q=20$). Figure \ref{FigSigvsEps}-(b) mimics the simulation of test 3, where the source a switched-on and off. But contrary to test 3, a nonlinear stress-strain relation is used. Large oscillations up to $t^*$ can be observed, contrary to what can be seen in figure \ref{FigTest3}.

\begin{figure}[h!]
\begin{center}
\begin{tabular}{cc}
(a) & (b)\\
\hspace{-0.5cm}
\includegraphics[scale=0.32]{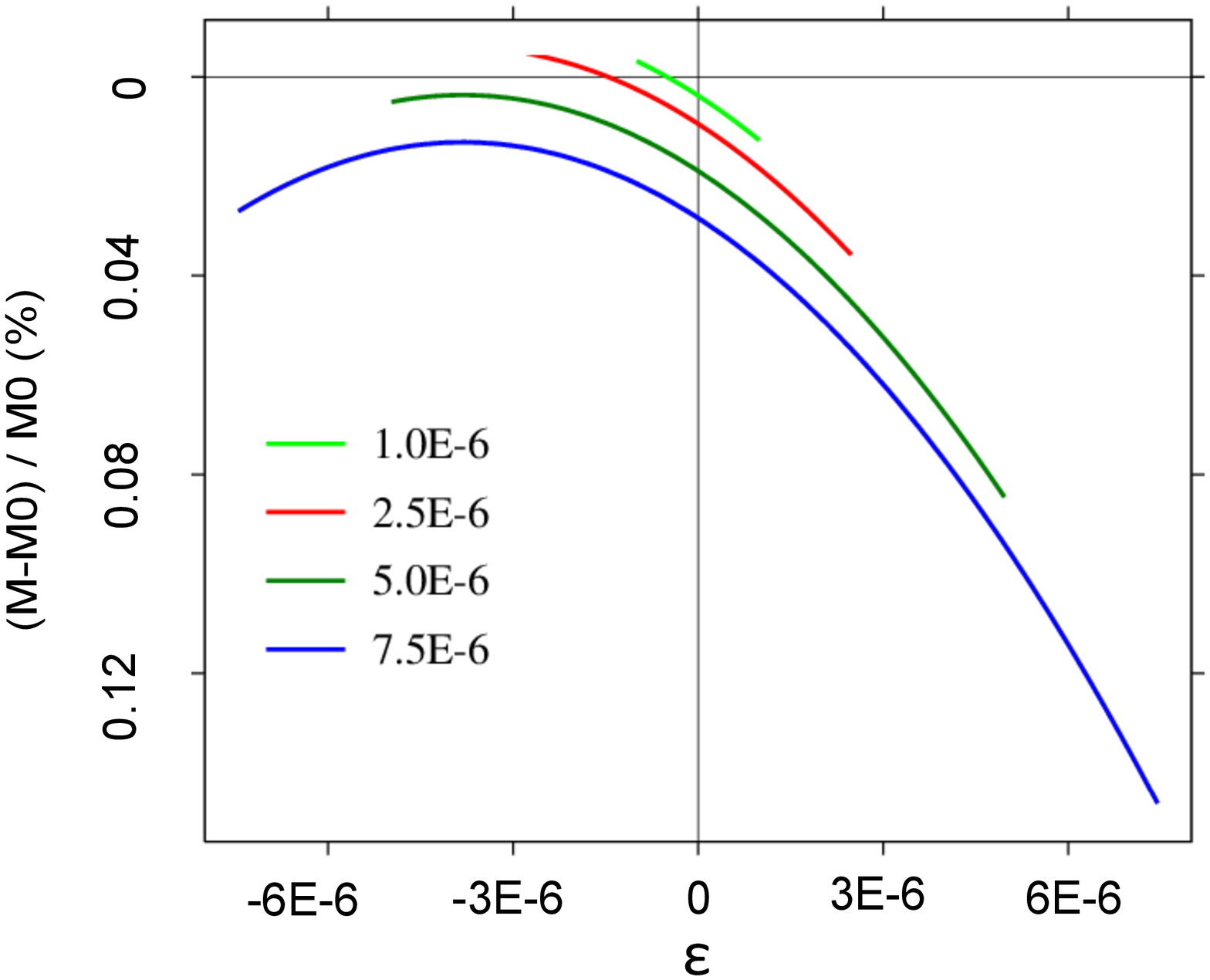} & 
\hspace{-0.9cm}
\includegraphics[scale=0.32]{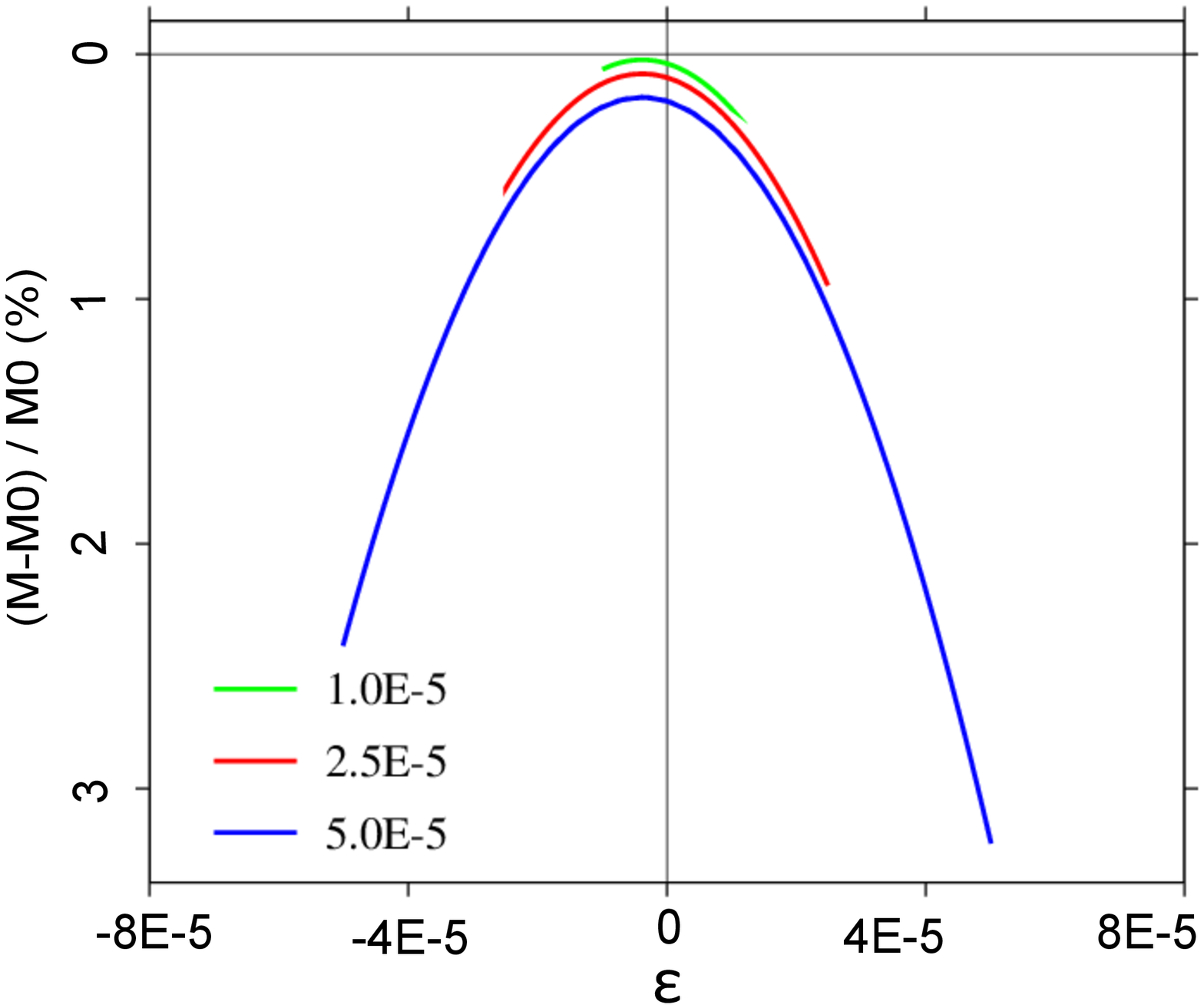} \\
(c) & (d)\\
\hspace{-0.5cm}
\includegraphics[scale=0.32]{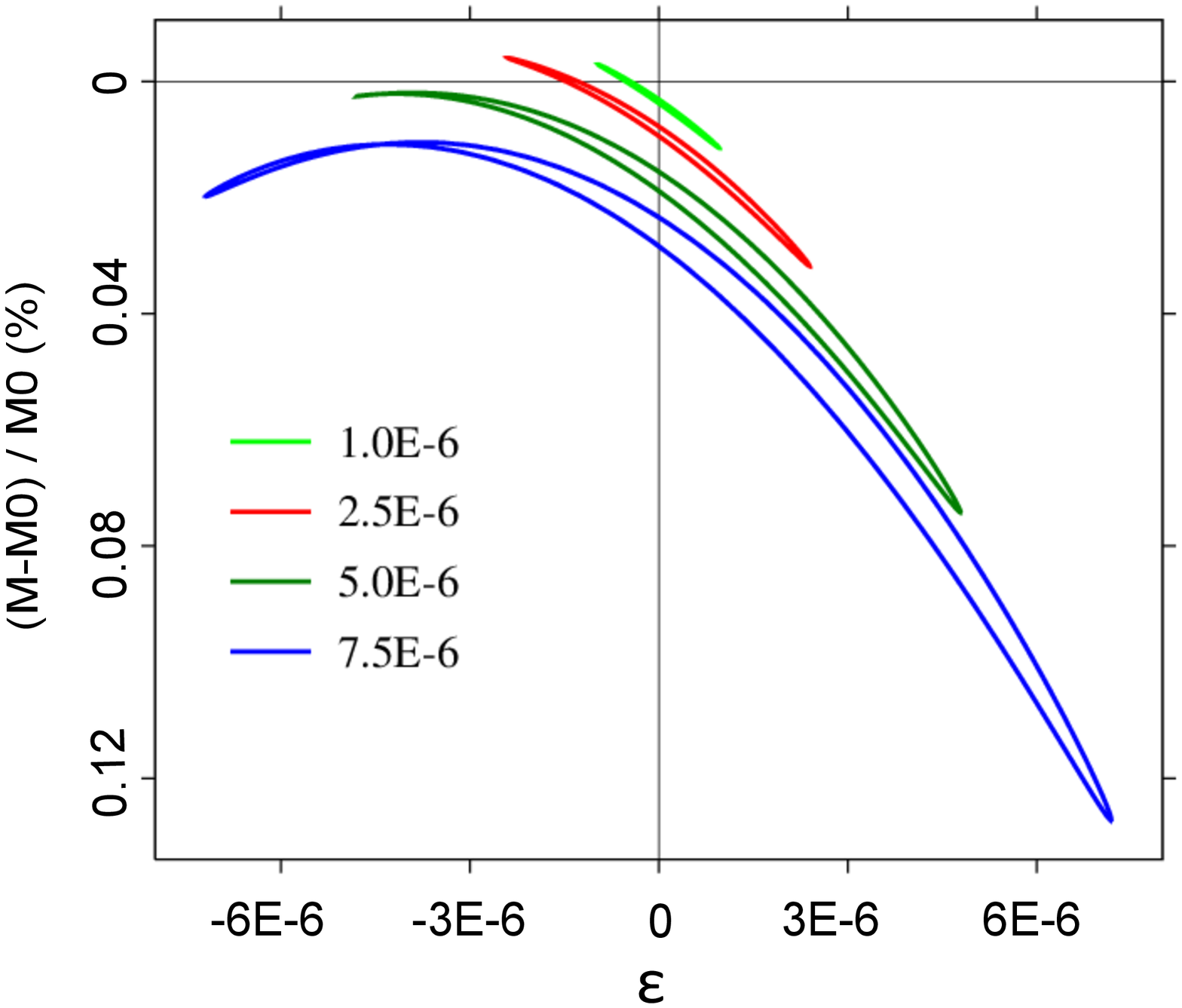} & 
\hspace{-0.9cm}
\includegraphics[scale=0.32]{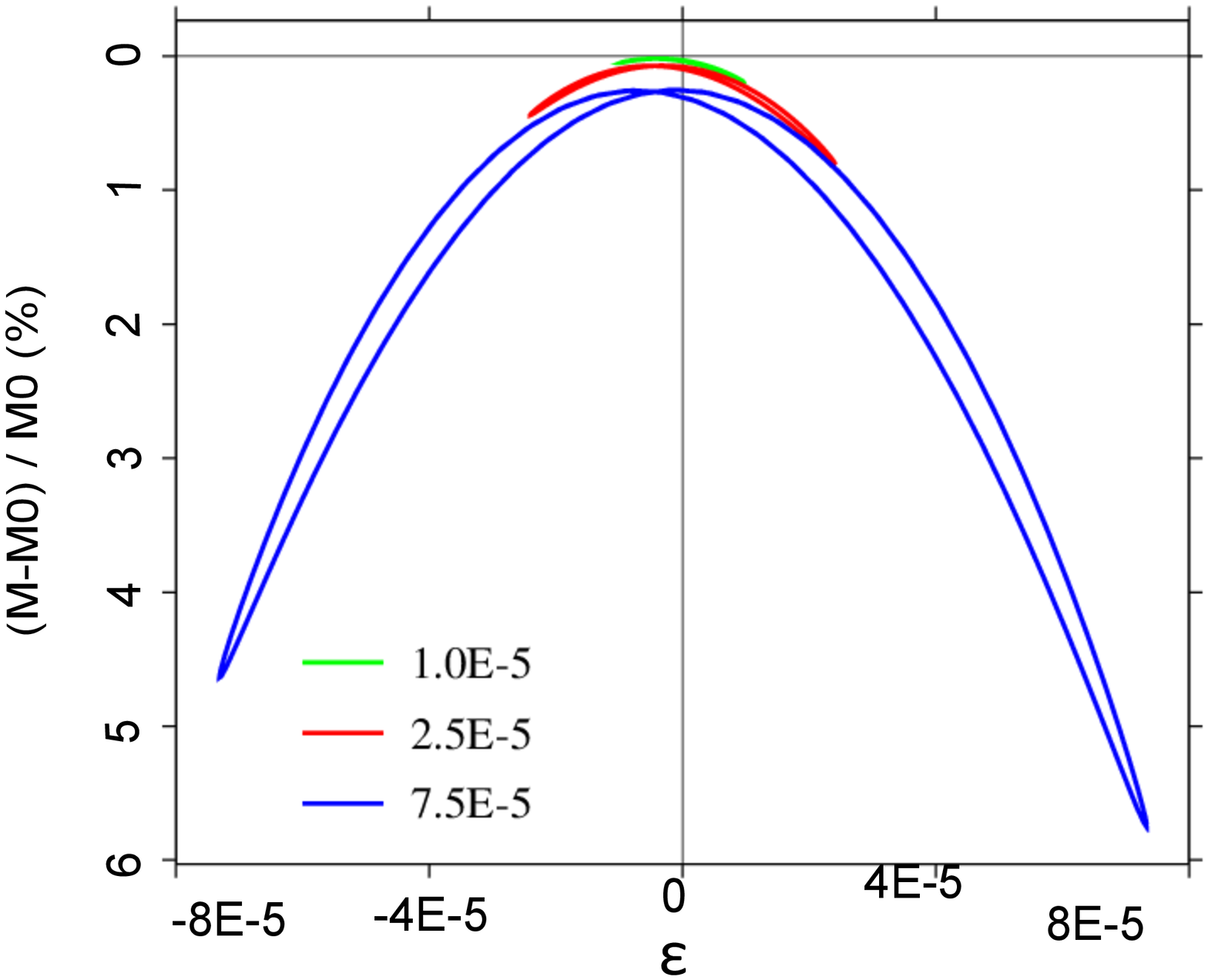} 
\end{tabular}
\caption{\label{FigMvsEps} test 4. Relative variations in the elastic modulus $M$ for various amplitudes of forcing $\varepsilon_{\max}$, from $10^{-6}$ to $7.5\,10^{-5}$. Top (a-b): without viscoelasticity; bottom (c-d): with viscoelasticity.}
\end{center}
\end{figure}

Figure \ref{FigMvsEps} displays the relative variation of the elastic modulus $\Delta M=(M-M_0)/M_0$ in terms of the strain, for various amplitudes of the forcing. Three observations can be made. First, nonlinear curves are obtained, which is a signature of the nonlinear stress-strain relation. Second, $\Delta M$ increases with $\varepsilon_{\max}$: softening increases monotonically with the forcing. Third and last, loops are obtained if and only if viscoelasticity is incorporated (c-d). These three features are qualitatively similar to those obtained experimentally  \cite{Renaud12,Riviere13}.


\subsection{Conclusion}\label{SecConclu}

We have proposed a one-dimensional model that captures the behavior of real media under longitudinal bar excitation, including the following features: nonlinear elasticity; softening / recovering of the elastic modulus; hysteretic evolution of the elastic modulus with the strain. The proposed model is very modular. It involves three different bricks which can be used also independently: see for instance the numerical experiments in section \ref{SecExp}, in which are considered various combinations of elasticity, attenuation and slow dynamics. Experimentally, the parameters corresponding to each mechanisms can be identified separately:
\begin{itemize}
\item the measure of nonlinear elastic parameters is described in many books \cite{Guyer99,Hamilton98};
\item the measure of the quality factor must be performed in the linear regime. See the reference book \cite{Bourbie87} for a description of an experimental protocol;
\item lastly, measuring the parameters of the slow dynamics is detailed in many papers cited in the bibliography. The current challenge is to link the physical observations to the parameters of Vakhnenko's model. Our ambition, with the present paper, is to provide experimenters with a tool for testing various sets of parameters, and hence testing the validity of Vakhnenko's model.
\end{itemize}
A major interest of the numerical approach is the possibility to tackle with variable coefficients in space, which is representative of localized defects \cite{Pecorari14}. In particular, a random initial distribution of defects $g_0(x)$ can be considered  straightforwardly.

Many improvements can be investigated, to mention but a few. More sophisticated models can be built quite naturally, considering for instance relaxation of the nonlinear coefficients ${\bf p}$ in (\ref{Rheo1}), or a nonlinear law in (\ref{Rheo2}). Concerning the numerical simulations, higher-order schemes (such as WENO schemes \cite{Leveque02}) can easily be adapted to the proposed formulation. Lastly, theoretical analyses should be done to prove rigorously the well-posedness of the model and its thermodynamic properties.

Work is currently proceeding along two directions. First, numerical simulations are being done to recover quantitatively the experimental results of the litterature \cite{Renaud12,Riviere13}. Second, the extension of this model to 2D and 3D geometries is under progress.


\appendix

\section{Parameters of the viscoelastic model}\label{SecViscoCoeff}

Standard calculations on (\ref{MailleSigma}), (\ref{Rheo}) and (\ref{TauEta}) yield the reciprocal of the quality factor $Q$ \cite{Carcione07}
\begin{equation}
Q^{-1}(\omega)=\left(\sum_{\ell=1}^N\frac{\omega\left(\tau_{\varepsilon\ell}-\tau_{\sigma\ell}\right)}{1+\omega^2\tau^2_{\sigma\ell}}\right)/\left(\displaystyle\sum_{\ell=1}^N\frac{1+\omega^2\tau_{\varepsilon\ell}\tau_{\sigma\ell}}{1+\omega^2\tau^2_{\sigma\ell}}\right).
\label{Qm1}
\end{equation}
Optimizing $Q^{-1}$ towards a given law (for instance a constant quality factor on a frequency range of interest $[f_{\min},\,f_{\max}]$) provides a means to determine $\tau_{\sigma\ell}$ and $\tau_{\varepsilon\ell}$ \cite{Lombard11}. Here an optimization with constraint is applied to ensure positive values of $\tau_{\sigma\ell}$ and $\tau_{\varepsilon\ell}$, as required by the decrease in energy (see section \ref{SecMathProp}). See \cite{BenJazia14} for details about such an optimization.

\begin{figure}[h!]
\begin{center}
\begin{tabular}{cc}
(a) & (b)\\
\hspace{-0.9cm}
\includegraphics[scale=0.35]{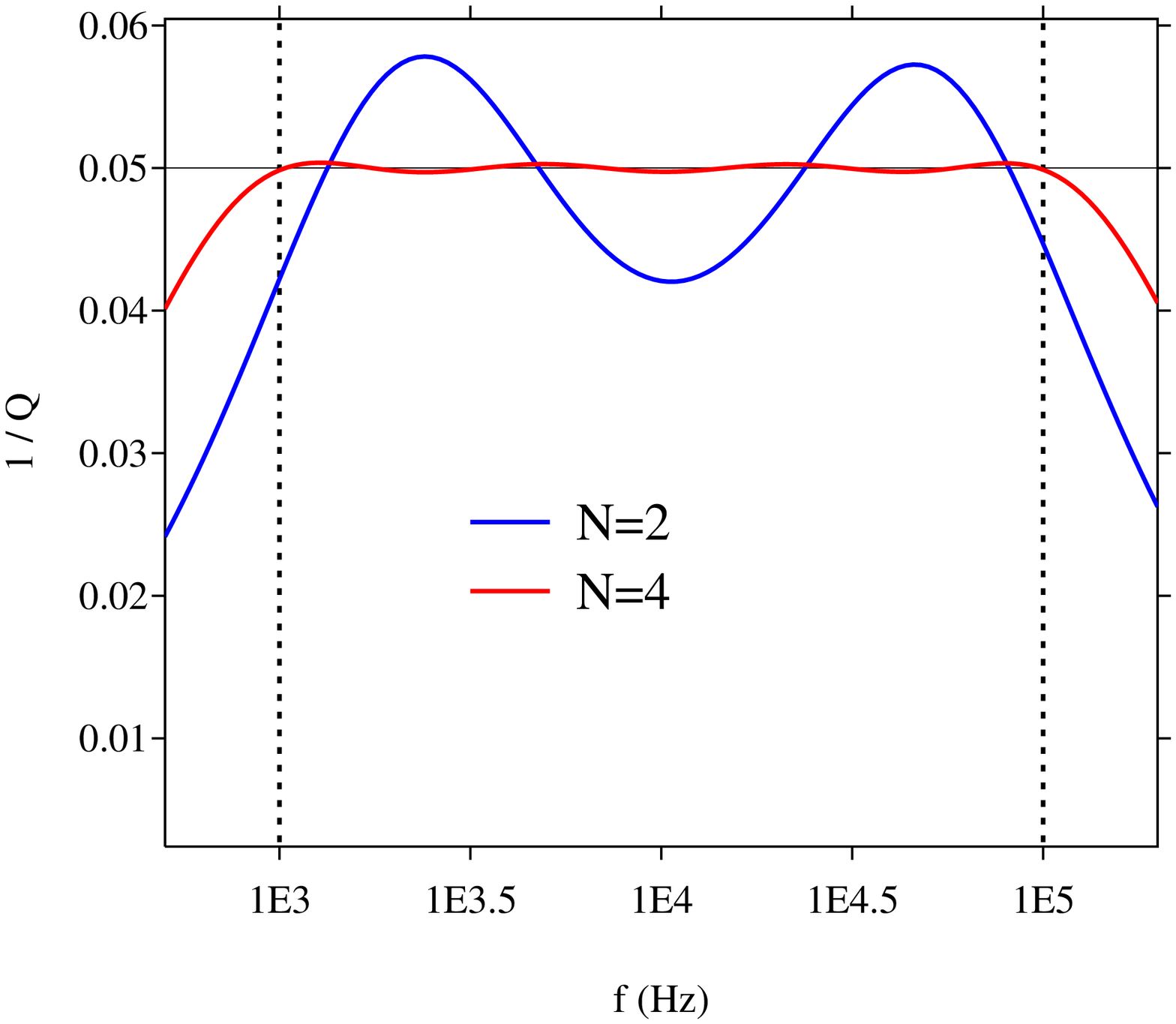} &
\hspace{-0.9cm}
\includegraphics[scale=0.35]{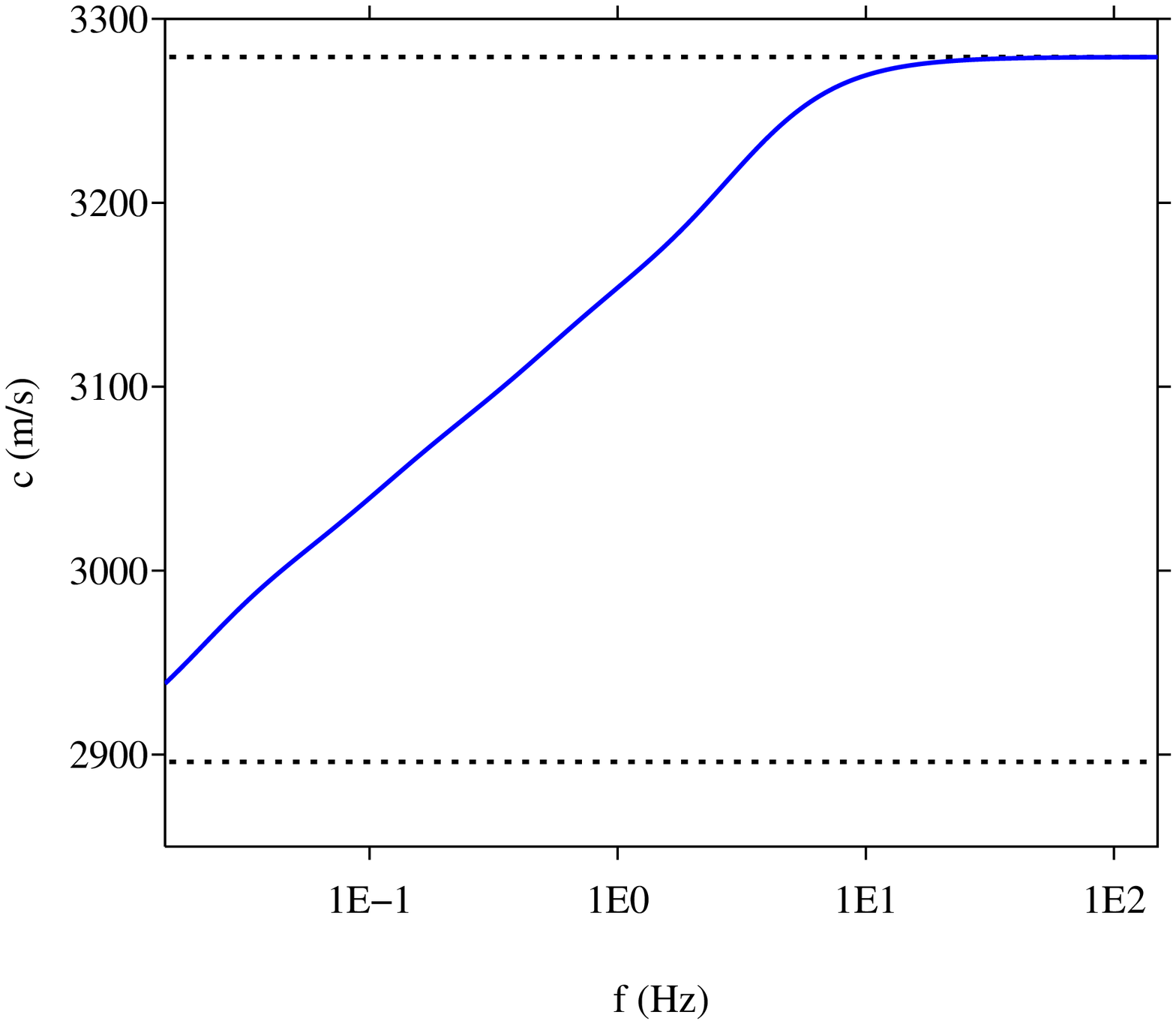} 
\end{tabular}
\caption{\label{FigPsiCel} Properties of the viscoelastic model in the linear regime. (a): reciprocal of the quality factor $Q=20$ (\ref{Qm1}). The constant exact value is denoted by a horizontal line; the values obtained after optimization with $N=2$ and $N=4$ relaxation mechanisms are denoted in blue and red, respectively; the range of optimization $[f_{\min},\,f_{\max}]$ is denoted by vertical dotted lines. (b): frequency evolution of the phase velocity; the horizontal dotted lines denote the phase velocity at zero and infinite frequency.}
\end{center}
\end{figure}

Figure \ref{FigPsiCel} illustrates the properties of the viscoelastic model. Figure \ref{FigPsiCel}-(a) compares the reciprocal of the constant quality factor $Q=20$ with the value deduced from (\ref{Qm1}), for $N=2$ and $N=4$ relaxation mechanisms. Nonlinear optimization is performed from $f_{\min}=1$ kHz to $f_{\max}=100$ kHz. Large oscillations are obtained for $N=2$; excellent agreement is observed for $N=4$. Figure \ref{FigPsiCel}-(b) shows the increase in phase velocity from $c_0=\sqrt{E_R/\rho}$ to $c_\infty=\sqrt{E/\rho}$. The reader is referred to \cite{Carcione07} for details about these quantities.

Lastly, the consistancy relation (\ref{Consistance}) is proven here.
Null attenuation amounts to an infinite quality factor. Equation (\ref{Qm1}) implies that $Q=+\infty$ is obtained if $\tau_{\varepsilon\ell}=\tau_{\sigma\ell}$. In this case, the viscoelastic coefficients (\ref{ErVsE}) and (\ref{E1E2Eta}) are
\begin{equation}
E_R=E,\hspace{0.5cm} K_{1\ell}=\frac{E}{N},\hspace{0.5cm} K_{2\ell}=+\infty,\hspace{0.5cm} \eta_\ell=+\infty.
\label{Qinfty}
\end{equation}
To get a bounded stress, (\ref{Rheo3}) implies $\varepsilon_{2\ell}=0$, and hence $\varepsilon_{1\ell}=\varepsilon$ for $\ell=1,\cdots,\,N$ (\ref{MailleEps}). Putting together the total stress (\ref{MailleSigma}), the nonlinear elasticity (\ref{ElastoNL}) and the homogeneity property in (\ref{PropElastoNL}), one obtains
\begin{equation}
\sigma=\sum_{\ell=1}^N s(\varepsilon_{1\ell},\,K_{1\ell},\,{\bf p})=\sum_{\ell=1}^N s\left(\varepsilon,\,\frac{E}{N},\,{\bf p}\right)=\frac{1}{N}\sum_{\ell=1}^Ns(\varepsilon,\,E,\,{\bf p})=s(\varepsilon,\,E,\,{\bf p}),
\label{ProofProp1}
\end{equation}
which concludes the proof.


\section{Analysis of hyperbolicity}\label{ProofHyp}

The Jacobian ${\bf A}$ of ${\bf f}$ (\ref{FluxFunc}) is 
\begin{equation}
{\bf A}({\bf U})=
\left(
\begin{array}{cccccc}
0      & 0      & \Phi_1  & \cdots & \Phi_N & 0\\
-1     & 0      & 0       & \cdots & 0      & 0\\
-1     & 0      & 0       & \cdots & 0      & 0\\
\vdots & \vdots & \vdots  &        & \vdots & \vdots  \\
-1     & 0      & 0       & \cdots & 0      & 0\\
0      & 0      & 0       & 0      & 0      & 0
\end{array}
\right),
\label{MatA}
\end{equation}
where
\begin{equation}
\Phi_\ell=-\frac{1}{\rho}\frac{\partial \sigma_{1\ell}}{\partial \varepsilon_{1\ell}}.
\label{Phi}
\end{equation}
The determinant of ${\bf A}$ writes
\begin{equation}
P_{\bf A}(\lambda)=-\lambda
\left|
\begin{array}{cccccc}
-\lambda &  0       & \Phi_1   & \cdots & \Phi_N  \\
-1       & -\lambda & 0        & \cdots & 0       \\
-1       & 0        & -\lambda &        & 0       \\
\vdots   &          &  \ddots  & \ddots &         \\
-1       &          &          &  0     & -\lambda
\end{array}
\right|
\end{equation}
The columns and lines are denoted by ${\cal C}_j$ and ${\cal L}_j$, respectively. The following algebraic manipulations are performed successively:
\begin{description}
\item[(i)] $\,{\cal C}_{1}\leftarrow \lambda\,{\cal C}_{1}$,
\item[(ii)] ${\cal C}_{1}\leftarrow {\cal C}_{1}-{\cal C}_j$, with $j=2,\,\cdots,\,N+1$,
\end{description}
which yields
\begin{equation}
\begin{array}{llll}
\lambda\,P_{\bf A}(\lambda)&=&-\lambda
\left|
\begin{array}{ccccc}
 \ds  -\lambda^2 - \sum_{\ell=1}^N\Phi_\ell &  0       & \Phi_1  & \cdots & \Phi_N   \\
0                                           & -\lambda &         &        &          \\
\vdots                                      &          &         & \ddots &          \\
0                                           &          &         &        & -\lambda
\end{array}
\right|,\\
&=& \displaystyle (-1)^{N+1}\lambda^{N+2}\left(\lambda^2+\sum_{\ell=1}^N\Phi_\ell\right).
\end{array}
\end{equation}
It follows that the eigenvalues are 0 (with multiplicity $N+1$) and $\pm c$, with the sound velocity (\ref{C2}). From (\ref{Phi}), real eigenvalues are obtained if and only if $c^2>0$ in  (\ref{C2}). 

Necessary and sufficient conditions are easily deduced from (\ref{C2}) for the models (\ref{Model1})-(\ref{Model3}) when $N=1$: hyperbolicity is satisfied if $|\varepsilon|<\varepsilon_c$, where
\begin{equation}
\varepsilon_c=
\left\{
\begin{array}{l}
\displaystyle
d\left(\left(\frac{\textstyle r+1}{\textstyle a+1}\right)^{\frac{\textstyle 1}{\textstyle r-a}}-1\right) \hspace{7cm}\mbox{(model 1)},\\
\\
+\infty \hspace{9.5cm}\mbox{(model 2)},\\
\\
\displaystyle \frac{\textstyle 1}{\textstyle 2\beta} \, \mbox{ if } \delta=0, \quad \frac{\textstyle \beta}{\textstyle 3\delta}\left(\sqrt{1+\frac{\textstyle 3\delta}{\textstyle \beta^2}}-1\right) \mbox{ otherwise} \hspace{3.65cm}\mbox{(model 3)}.
\end{array}
\right.
\label{EpsC}
\end{equation}
Model 2 is always hyperbolic. On the contrary, the widely-used Landau model (model 3) is conditionally hyperbolic. When $N>1$, the hyperbolicity condition $|\varepsilon_{1\ell}|<\varepsilon_c$ is sufficient.

Given the nonlinear elastic models (\ref{Model1})-(\ref{Model3}), the speed of sound $c$ satisfies:
\begin{equation}
c^2=
\left\{
\begin{array}{l}
\displaystyle
\sum_{\ell=1}^N\frac{\textstyle K_{1\ell}}{\textstyle \rho}\,\frac{\textstyle 1}{\textstyle r-a}\left(\ds \frac{\textstyle r+1}{\textstyle \left(\ds 1+\frac{\textstyle \varepsilon_{1\ell}}{\textstyle d}\right)^{r}}-\frac{\textstyle a+1}{\textstyle \left(\ds 1+\frac{\textstyle \varepsilon_{1\ell}}{\textstyle d}\right)^{a}}\right)   \hspace{4.53cm} \mbox{(model 1)},\\
\\
\displaystyle
\sum_{\ell=1}^N\frac{\textstyle K_{1\ell}}{\textstyle \rho}\,\left(1-(r+a+3)\frac{\textstyle \varepsilon_{1\ell}}{\textstyle d}+\frac{\textstyle 1}{\textstyle 2}\left(r^2+ra+a^2+6r+6a+11\right)\left(\frac{\textstyle \varepsilon_{1\ell}}{\textstyle d}\right)^2\right) \hspace{0.3cm} \mbox{(model 2)},\\
\\
\displaystyle
\sum_{\ell=1}^N\frac{\textstyle K_{1\ell}}{\textstyle \rho}\,\left(1-2\,\beta\,\varepsilon_{1\ell}-3\,\delta\,\varepsilon_{1\ell}^2\right) \hspace{6.00cm} \mbox{(model 3)}.
\end{array}
\right.
\label{C2Modeles}
\end{equation}


\section{Analysis of the relaxation terms}\label{ProofRelax}

For linear stress-strain relations (\ref{Rheo}), the relaxation coefficients (\ref{DeltaL}) yield 
\begin{equation}
\left\{
\begin{array}{l}
\displaystyle
\frac{\partial \Delta_\ell}{\partial \varepsilon}(0)=\frac{1}{\eta_\ell}\sigma^{'}_{2\ell}(0)=\frac{K_{2\ell}}{\eta_\ell},\\
[8pt]
\displaystyle
\frac{\partial \Delta_\ell}{\partial \varepsilon_{1\ell}}(0)=-\frac{1}{\eta_\ell}\left(\sigma^{'}_{1\ell}(0)+\sigma^{'}_{2\ell}(0)\right)=-\frac{1}{\eta_\ell}\left(K_{1\ell}+K_{2\ell}\right).
\end{array}
\right.
\end{equation}
The Jacobian matrix of the relaxation function (\ref{FluxFunc}) can be obtained
\begin{equation}
{\bf J}=
\left(
\begin{array}{ccccc}
0                         & 0                                 & \cdots & 0  & 0                               \\
\ds \frac{E_{21}}{\eta_1} & \ds -\frac{E_{11}+E_{21}}{\eta_1} &        &    & 0                               \\
\vdots                    &                                   & \ddots &    &                                    \\
\ds \frac{E_{2N}}{\eta_1} &                                   &        & \ds -\frac{E_{1N}+E_{2N}}{\eta_N} & 0\\
0 &  & \cdots & 0 & f_\xi
\end{array}
\right),
\label{MatJ}
\end{equation}
with $f_\xi=f_r$ if $g>g_\sigma$, $f_\xi=f_d$ if $g<g_\sigma$, $f_\xi=0$ else. It follows that the eigenvalues are 0, $-\frac{K_{1\ell}+K_{2\ell}}{\eta_\ell}$, and $-f_\xi$. 


\section{Semi-analytical solution}\label{SecSolExact}

The semi-analytical solution of the viscodynamic equations is computed as follows. Fourier transforms in space and time are applied to the system (\ref{EDPVisco}). Applying an inverse Fourier transform in space yields
\begin{equation}
{\hat v}(x,\omega)=\frac{i\omega\,\rho}{\ds \sum_{\ell=1}^NK_{1\ell}\frac{i\omega+1/\tau_{\varepsilon_\ell}}{i\omega+1/\tau_{\sigma_\ell}}}\frac{\hat{\cal G}(\omega)}{2\,\pi}\int_{-\infty}^{+\infty}\frac{1}{k^2-k_0^2}e^{-ikx_0}\,dk,
\label{Fourier}
\end{equation}
where the hat refers to the Fourier transform, ${\cal G}$ is the time evolution of the source, the relaxation times $\tau_{\varepsilon_\ell}$ and $\tau_{\sigma_\ell}$ are defined in (\ref{TauEta}), and $k$ is the wavenumber. The poles $\pm k_0$ satisfy
\begin{equation}
k_0^2=\frac{\rho\,\omega^2}{\ds \sum_{\ell=1}^NK_{1\ell}\frac{i\omega+1/\tau_{\varepsilon_\ell}}{i\omega+1/\tau_{\sigma_\ell}}}
\label{K0}
\end{equation}
with $\Im\mbox{m}(k_0)<0$. Applying the residue theorem gives the time-domain velocity
\begin{equation}
v(x,t)=\rho\int_0^\infty \Re\mbox{e}\left(\frac{\omega}{k_0}\frac{1}{\ds \sum_{\ell=1}^NK_{1\ell}\frac{i\omega+1/\tau_{\varepsilon_\ell}}{i\omega+1/\tau_{\sigma_\ell}}} e^{-ik_0|x-x_0|}\,\hat{\cal G}(\omega)\right)\,d\omega.
\label{FourierInv}
\end{equation}
Expressions for $\varepsilon$ and $\varepsilon_{1\ell}$ can be obtained in a similar manner. Lastly, the numerical evaluation of (\ref{FourierInv}) is done using a rectangular quadrature rule on $N_f$ Fourier modes and with a constant frequency step $\Delta f$ on the frequency band of interest.

\vspace{1cm}


\end{document}